\documentclass[preprint,3p,times,authoryear]{elsarticle}

\usepackage{graphicx}

\usepackage{amssymb, amsmath}

\usepackage{bm,fix-cm}

\usepackage{algorithm, setspace, booktabs}
\usepackage[noend]{algpseudocode}

\usepackage{bm,fix-cm}
\DeclareMathAlphabet{\mathsfit}{T1}{\sfdefault}{\mddefault}{\updefault}
\SetMathAlphabet{\mathsfit}{bold}{T1}{\sfdefault}{\bfdefault}{\updefault}

\usepackage{upgreek}
\newcommand{\vect}[1]{{\bm{\mathit{#1}}}}
\newcommand{\matr}[1]{\ensuremath{\boldsymbol{\mathsf{#1}}}}

\newcommand{\totald}{\ensuremath{\mathrm{d}}\,}
\newcommand{\T}{^\mathrm{T}}
\newcommand{\E}[1]{\ensuremath{\cdot 10^{#1}}}

\usepackage[dvipsnames]{xcolor}
\journal{}

\usepackage{hyperref}
\hypersetup{
    colorlinks=true,
    linkcolor={red!80!black},
    citecolor={red!80!black},
    urlcolor={red!80!black}
}

\begin{document}
\begin{frontmatter}

\title{A real-time digital twin of azimuthal thermoacoustic instabilities}
\author[inst1,inst2]{Andrea Nóvoa}
\author[inst3]{Nicolas Noiray}
\author[inst4]{James R. Dawson}
\author[inst2,inst5,inst6,inst1]{Luca Magri}

\affiliation[inst1]{organization={Cambridge University Engineering Dept.},
            city={Cambridge},
            postcode={CB2 1PZ},
            country={UK}}

\affiliation[inst2]{organization={Imperial College London, Aeronautics Dept.},
            city={London},
            postcode={SW7 2AZ}, 
            country={UK}}

\affiliation[inst3]{
organization={ETH Zürich, Mechanical \& Process Engineering Dept.},
city={Zürich},
postcode={8092},
country={Switzerland}
}
\affiliation[inst4]{
organization={Norwegian University of Science and Technology 
(NTNU), Dept. of Energy and Process Engineering},
city={Trondheim},
country={Norway}
}

\affiliation[inst5]{organization={The Alan Turing Institute},
            addressline={96 Euston Rd}, 
            city={London},
            postcode={NW1 2DB}, 
            country={UK}}

\affiliation[inst6]{organization={Dipartimento di Ingegneria Meccanica e Aerospaziale, Politecnico di Torino},
city={Torino},
postcode={10129},
country={Italy}
} 

\begin{abstract}

When they occur, azimuthal thermoacoustic oscillations can detrimentally affect the safe operation of gas turbines and aeroengines. 
We develop a real-time digital twin of azimuthal thermoacoustics of a hydrogen-based annular combustor. 
The digital twin seamlessly combines two sources of information about the system 
(i) a physics-based low-order model; 
and (ii) raw and sparse experimental data from microphones, which contain both aleatoric noise and turbulent fluctuations. First, we derive a low-order thermoacoustic model for azimuthal instabilities, which is deterministic. Second, we propose a real-time data assimilation framework to infer the acoustic pressure, the physical parameters, and the model bias and measurement shift simultaneously. This is the bias-regularized ensemble Kalman filter (r-EnKF), for which we find an analytical solution that solves the optimization problem. Third, we propose a reservoir computer, which infers both the model bias and measurement shift to close the assimilation equations. 
Fourth, we propose a real-time digital twin of the azimuthal thermoacoustic dynamics of a laboratory hydrogen-based annular combustor for a variety of equivalence ratios. 
 We find that the real-time digital twin
(i) autonomously predicts azimuthal dynamics, in contrast to bias-unregularized methods; 
(ii) uncovers the physical acoustic pressure from the raw data, i.e., it acts as a physics-based filter;
(iii) is a time-varying parameter system, which generalizes existing models that have constant parameters, and capture only slow-varying variables.
The digital twin generalizes to all equivalence ratios, which bridges the gap of existing models. This work opens new opportunities for real-time digital twinning of multi-physics problems. 

\end{abstract}

\end{frontmatter}

\section{Introduction}
\label{sec:intro}

Thermoacoustic instabilities are a multi-physics phenomenon, which is caused by the constructive coupling between hydrodynamics, unsteady heat released by flames, and acoustics~\citep[e.g.,][]{paschereit_structure_1998, culick_combustion_1988,  lieuwen_mechanism_2001, candel_flame_2009, poinsot_prediction_2017, juniper_sensitivity_2018, silva_intrinsic_2023, magri_ARFM_2023}. 
A thermoacoustic instability can arise when the heat released by the flames is sufficiently in phase
with the acoustic pressure \citep{rayleigh_explanation_1878, magri_sensitivity_2020}. 
If uncontrolled or not prevented, instabilities grow into large-amplitude pressure oscillations,
which can detrimentally affect gas turbine operating regimes, cause structural damage and fatigue, and, in the worst-case
scenario, shake the engine and its components apart \citep[e.g.,][]{candel_combustion_2002, dowling_feedback_2005, culick_unsteady_2006, lieuwen_book_2012}.
In aeroengines, the flame holders are arranged in annular configurations to increase the
power density~\citep[e.g.,][]{krebs_thermoacoustic_2002}. 
Despite these annular configurations being nominally rotationally symmetric, large-amplitude azimuthal thermoacoustic oscillations can spontaneously occur and break dynamical symmetry~\citep[e.g.,][]{morgans_model_2007, noiray_investigation_2011, noiray_dynamic_2013,
bauerheim_poinsot_progress_2016, 
moek_orchini_symmetry_2023, magri_ARFM_2023}.
Thermoacoustic instabilities in annular combustors have intricate dynamics, which can be grouped into~\citep[e.g.,][]{magri_ARFM_2023} 
(a) spinning, if the nodal lines rotate azimuthally (typical of rotationally symmetric
configurations); 
(b) standing, if the nodal lines have, on average, a fixed orientation 
(typical of rotationally asymmetric configurations); 
and 
(c) mixed, if the nodal lines switch
between the two former states (typical of weakly asymmetric configurations)~\citep[e.g.,][]{schuermans_nonlinear_2006, noiray_investigation_2011,worth_modal_2013}. 
Azimuthal instabilities have been investigated by high-fidelity simulations~\citep[e.g.,][]{wolf_acoustic_2012}; by experimental campaigns in atmospheric and pressurized rigs  
\citep[e.g.,][]{bourgouin_candel_self_2013,worth_self_2013,ahn_worth_heat_2022, mazur_worth_characteristics_2019,mazur_worth_self_2021,indlekofer_noiray_spontaneous_2022} and heavy-duty gas turbines~\citep{noiray_dynamic_2013}; and  by 
theoretical studies~\citep[e.g.,][]{bauerheim_poinsot_progress_2016,ghirardo_azimuthal_2013, moeck_thermoacoustic_2010, noiray_investigation_2011, mensah_effects_2019,duran_reflection_2015,laera_flame_2017, murthy_analysis_2019,faure_noiray_symmetry_2020}.
Because the azimuthal dynamics are not yet fully understood~\citep{aguilar_locking_2021,faure_noiray_imperfect_2021}, 
the understanding, modelling, and control of azimuthal oscillations is an active area of research. \\ 
%

Experimental campaigns were performed to gain insight into the physical mechanisms and behaviour of azimuthal instabilities in a prototypical annular combustor with electrically heated gauzes~\citep{moeck_thermoacoustic_2010}.
In a model annular gas turbine combustor with ethylene-air flames, \citet{worth_self_2013} investigated the flame dynamics and heat release, and how it coupled with the acoustics~\citep{worth_modal_2013,oconnor_worth_flame_2013}. 
They found that varying the burner spacing, to promote or suppress flame-flame interactions, resulted in changes to the amplitude and frequency of the azimuthal modes. They also varied the burner swirl directions, which impacted the preferred mode selection.
The spontaneous symmetry breaking of thermoacoustic eigenmodes was experimentally analysed in~\citet{indlekofer_noiray_spontaneous_2022}, who found that small imperfections in the rotational symmetry of the annular combustor were magnified at the supercritical Hopf bifurcation point which separated resonant from limit cycle oscillations in the form of a standing mode oriented at an azimuthal angle defined by the rotational asymmetry. 
The nonlinear dynamics of beating modes was experimentally discovered and modelled by~\citet{faure_noiray_imperfect_2021}. 
The authors found that for some combinations of small asymmetries of the resistive and reactive components of the thermoacoustic system, purely spinning limit cycles became unstable, and heteroclinic orbits between the corresponding saddle points led to beating oscillations with periodic changes in the spin direction. 
The previous works focused on statistically stationary regimes. 
The analysis of slowly varying operating conditions were investigated in~\citet{indlekofer_noiray_effect_2021}, who unravelled dynamic hysteresis of the thermoacoustic state of the system. 
Other experimental and theoretical investigations focused on analysing the intermittent behaviour~\citep{roy_sujith_flame_2021,faure_noiray_experiments_2021}. In the latter work, the solutions of the Fokker-Plank equation, which governs the probability of observing an instantaneous limit cycle state dominated by a standing or spinning mode, enabled the prediction of the first passage time statistics between erratic changes in spin direction. \\

The intricate linear and nonlinear dynamics of azimuthal thermoacoustic oscillations spurred interest in physics-based low-order modelling and control~\citep[e.g.,][]{morgans_model_2007,illingworth_morgans_adaptive_2010,humbert_symmetry_2023}. 
The azimuthal dynamics can be qualitatively described by a one-dimensional wave-like equation~\citep{noiray_investigation_2011,ghirardo_azimuthal_2013,bauerheim_analytical_2014,bothien_noiray_analysis_2015,yang_morgans_systematic_2019},  
which can also include a stochastic forcing term to model the effect of the turbulent fluctuations and noise~\citep{noiray_deterministic_2013,orchini_degenerate_2020}. 
{In the frequency domain, azimuthal oscillations were investigated with eigenvalue sensitivity   \citet{magri_stabilityI_2016,magri_stabilityII_2016,mensah_effects_2019,orchini_degenerate_2020}, who found that traditional eigenvalue sensitivity analysis needed to be extended to tackle degenerate pairs of azimuthal modes, as reviewed in~\citet{ magri_ARFM_2023}. 
A review of azimuthal thermoacoustic modelling can be found in~\citet{bauerheim_analytical_2014}.
In the time domain, the typical approach is to develop models that describe the acoustic state, which is also referred to as the slow-varying variable approach based on quaternions~\citep[e.g.,][]{ghirardo_quaternion_2018} or as the generalized Bloch sphere representation~\citep{magri_ARFM_2023}. In this approach, all fast-varying dynamics, such as aleatoric noise and turbulent fluctuations, are filtered out of the data and modelled as a stochastic forcing term. The equations model the amplitude of the acoustic pressure envelope, the temporal phase drift, and the angles defining rotational and reflectional symmetry breaking~\citep{ghirardo_quaternion_2018}. \citet{faure_noiray_symmetry_2020} introduced the slow-varying variables into an stochastic wave equation, which was averaged in space and time to obtain coupled Langevin equations. The low-order model parameters were estimated via Langevin regression, which is a regression method used in turbulent environments~\citep[e.g.,][]{siegert_1998_analysis,noiray_linear_2017,boujo_processing_2020, callaham_nonlinear_2021}. 
Slow-varying variable models provide qualitatively accurate representations of azimuthal oscillations. However, they are not suitable for real-time applications, in which we need models that can receive data from sensors as raw inputs. 
Therefore, this work focus on models describing the fast-varying quantities, which  model the evolution of the modal amplitudes through coupled Van der Pol oscillators~\citep[e.g.,][]{noiray_investigation_2011}.
%
Nonetheless, the discussed low-order modelling methods  are offline, i.e., they infer the model parameters from the data in a post-processing stage and they identify one parameter at a time
(i.e., they do not estimate all the parameters in one computation). 
This means that the physical parameters of the literature are not necessarily optimal.  
\\

%
On the one hand, physics-based low-order models are qualitatively accurate, but they are quantitatively inaccurate due to modelling assumptions and approximations~\citep{magri_doan_2019}. The low-order model's state and parameters are affected by both aleatoric uncertainties and model biases~\citep{novoa_magri_inferring_2024}. On the other hand, experimental data can provide reliable information about the system, but they are typically noisy and sparse~\citep[e.g.,][]{magri_doan_2019}.  Experimental measurements can be affected by aleatoric uncertainties due to environmental and instrumental noise, and systematic errors in the sensors. To rigorously combine the two sources of information (low-order modelling and experimental data) to improve the knowledge on the system, data assimilation comes into play~\citep[e.g.,][]{tarantola2005inverse, evensen_book_2009}. 
Data assimilation (DA) for thermoacoustics was introduced with a variational approach (offline) by~\citet{traverso_data_2019}, and a real-time approach (sequential) by \citet{novoa_magri_2022}, who deployed an ensemble square-root Kalman filter to infer the pressure and physical parameters. Sequential data assimilation  has also been successfully applied to reacting flows~\citep[e.g.,][]{yu_data_2019, labahn_data_2019, donato_self_2024}, turbulence modelling~\citep[e.g.,][]{colburn_state_2011, majda_filtering_2012, gao_data_2017, magri_doan_2019, hansen_normal_2024}; and acoustics~\citep[e.g.,][]{kolouri_acoustic_2013, wang_data_2021}, among others. 
Apart from~\citet{novoa_magri_2022}, these works implemented classical DA formulations, which assume that the uncertainties are aleatoric and unbiased~\citep[e.g.,][]{dee_bias_2005,laloyaux_towards_2020}.  
However, as shown in \citet{novoa_magri_inferring_2024}, a DA framework provides an optimal state and set of parameters when the model biases are modelled.
Model bias estimation in real-time DA is traditionally based on the separate-bias Kalman filter scheme~\citep{friedland_treatment_1969}. 
This bias-aware filter augments the dynamical system with a parameterized model for the bias, and then solves two state and parameter estimation problems: one for the physical model, and another for the bias model~\citep[e.g.,][]{ignagni_separate_1990, dee_data_1998, dasilva_colonius_flow_2020}. 
However, the separate-bias Kalman filter relies on the knowledge of the bias functional form {\it a priori}. 
Also, as explained in~\citet{novoa_magri_inferring_2024}, the filter is unregularized, which can lead to unrealistically large estimates of the model bias and does not ensure that the bias is unique. In other words, the separate-bias Kalman filter may be ill-posed. 
To overcome these limitations, \citet{novoa_magri_inferring_2024} derived the bias-aware regularized ensemble Kalman filter (r-EnKF), for which an analytical solution was found. 
The r-EnKF is a real-time DA method that not only accounts for the model bias, but also regularizes the norm of the bias, which makes the algorithm stable and the solution of the problem unique. 
The model bias was inferred by an Echo State Network (ESN), which is a universal approximator of time-varying functions~\citep[e.g.,][]{GRIGORYEVA2018495} that 
can adaptively estimate the model bias with no major assumption regarding the functional form~\citep{novoa_magri_inferring_2024}. 
ESNs are suitable for real-time digital twins because their training consists of solving a linear system, which is computationally inexpensive in contrast to the  back-propagation methods required in other machine learning methods~\citep[e.g.,][]{bonavita_machine_2020,brajard_boquet_combining_2020}. 
The bias-regularized EnKF of~\citet{novoa_magri_inferring_2024} assumed that the measurements were unbiased, which is an assumption that we will relax in this paper to deal with actual experimental data. 

\subsection{Objectives and structure}

The objective of this work is to develop a real-time digital twin of azimuthal thermoacoustic instabilities.
Real-time digital twins are adaptive models, which are designed to predict the behaviour of their physical counterpart by assimilating data from sensors when it becomes available. 
We need four ingredients to design a real-time digital twin 
(i) data from sensors (from noisy and sparse microphones); 
(ii) a qualitative low-order model, which should capture the fast-varying acoustic dynamics rather than the slow-varying states; 
(iii) an estimator of both the model bias and measurement shift, 
(iv) a statistical data assimilation method that optimally combines the data and the model on the fly (real time) to infer the physical states and parameters. 
\S\ref{sec:azimuthal_TA} describes the available experimental data, as well as the physics-based low-order model used to describe thermoacoustic problem.
The real-time assimilation framework is detailed in \S\ref{sec:data_assimilation}. 
The proposed ESN is described in \S\ref{sec:ESN}. 
\S\ref{sec:results} shows the results of the real-time digital twin, and compares the performance of the bias-unregularized EnKF  with the bias-regularized r-EnKF. 
\S\ref{sec:conclusions} ends the paper.

\section{Azimuthal thermoacoustic instabilities}\label{sec:azimuthal_TA}

We model in real time azimuthal thermoacoustic instabilities in hydrogen-based annular combustors.  In this section, we describe both the experimental data (\S\ref{sec:exp_setup}) and the physics-based low-order model (\S\ref{sec:LOM_annular}).

\subsection{Experimental setup and data}\label{sec:exp_setup}

\begin{figure}
    \centering
    \includegraphics[width=.6\textwidth]{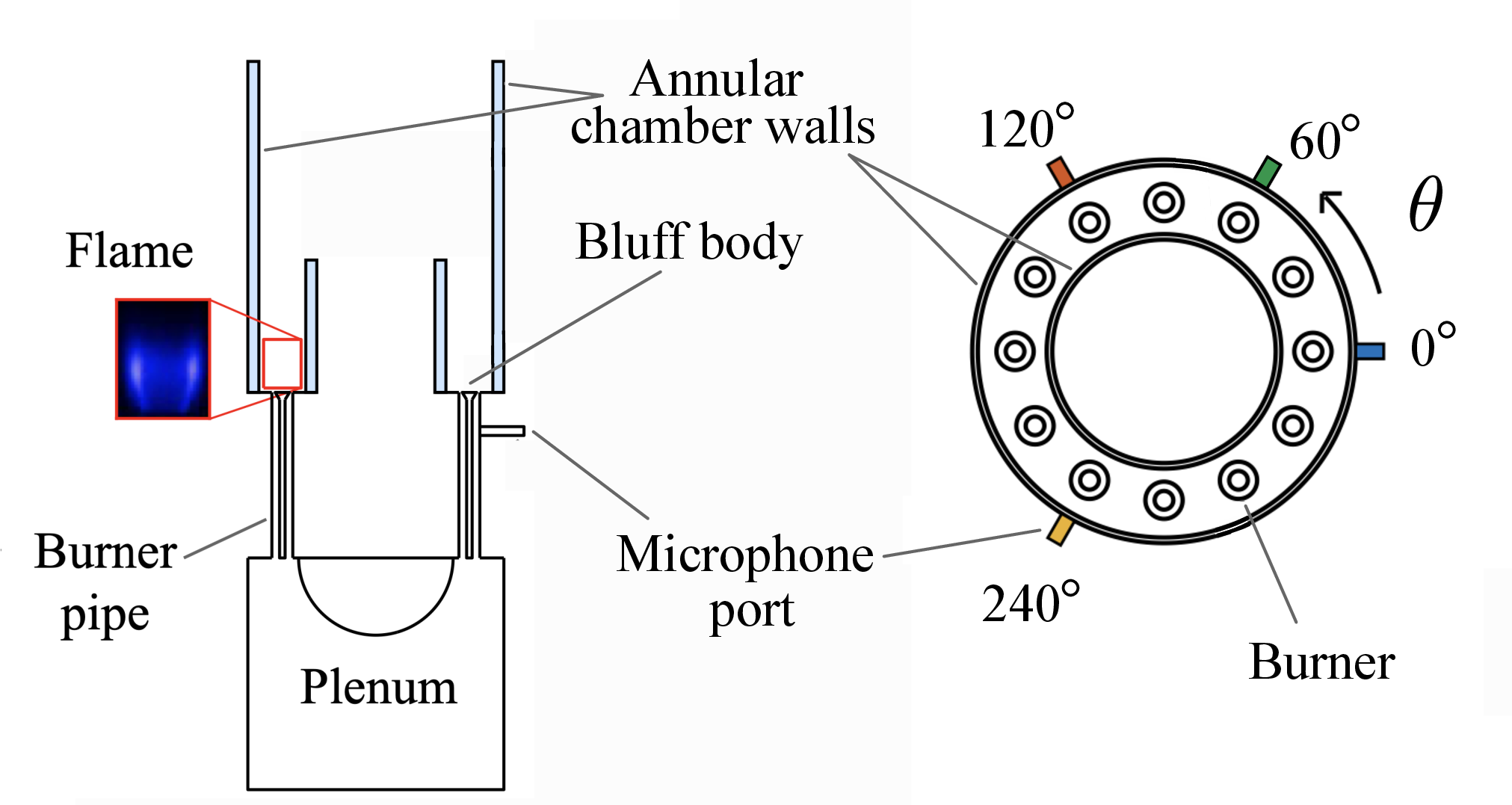}
    \caption{Side and top views of the experimental setup of the annular combustor. Adapted from \cite{faure_noiray_imperfect_2021}.}
    \label{fig:experimental_setup}
\end{figure}

We use the experimental data of \citet{faure_noiray_experiments_2021} and \citet{indlekofer_noiray_spontaneous_2022}. 
The experimental setup (Figure~\ref{fig:experimental_setup}) consists of  an annular combustor with twelve equally-spaced burners with premixed flames, which are fuelled with a mixture of 70/30\%  H$_2$/CH$_4$ by power. The operating conditions are atmospheric, and the thermal power is fixed at 72~kW.  
The equivalence ratios considered are $\Phi=\{0.4875, 0.5125, 0.5375, 0.5625\}$. When $\Phi >0.5$ the system is thermoacoustically unstable with  self-sustained oscillations  that peak at approximately 1090~Hz~\citep{indlekofer_noiray_spontaneous_2022}. 
The acoustic pressure data was recorded at a sampling rate of  51.2~kHz  by Kulite  pressure transducers (XCS-093-05D) at four azimuthal locations $\theta=\{0^\circ, 60^\circ, 120^\circ, 240^\circ\}$.  

\begin{figure}
    \centering
    \includegraphics[width=\textwidth]{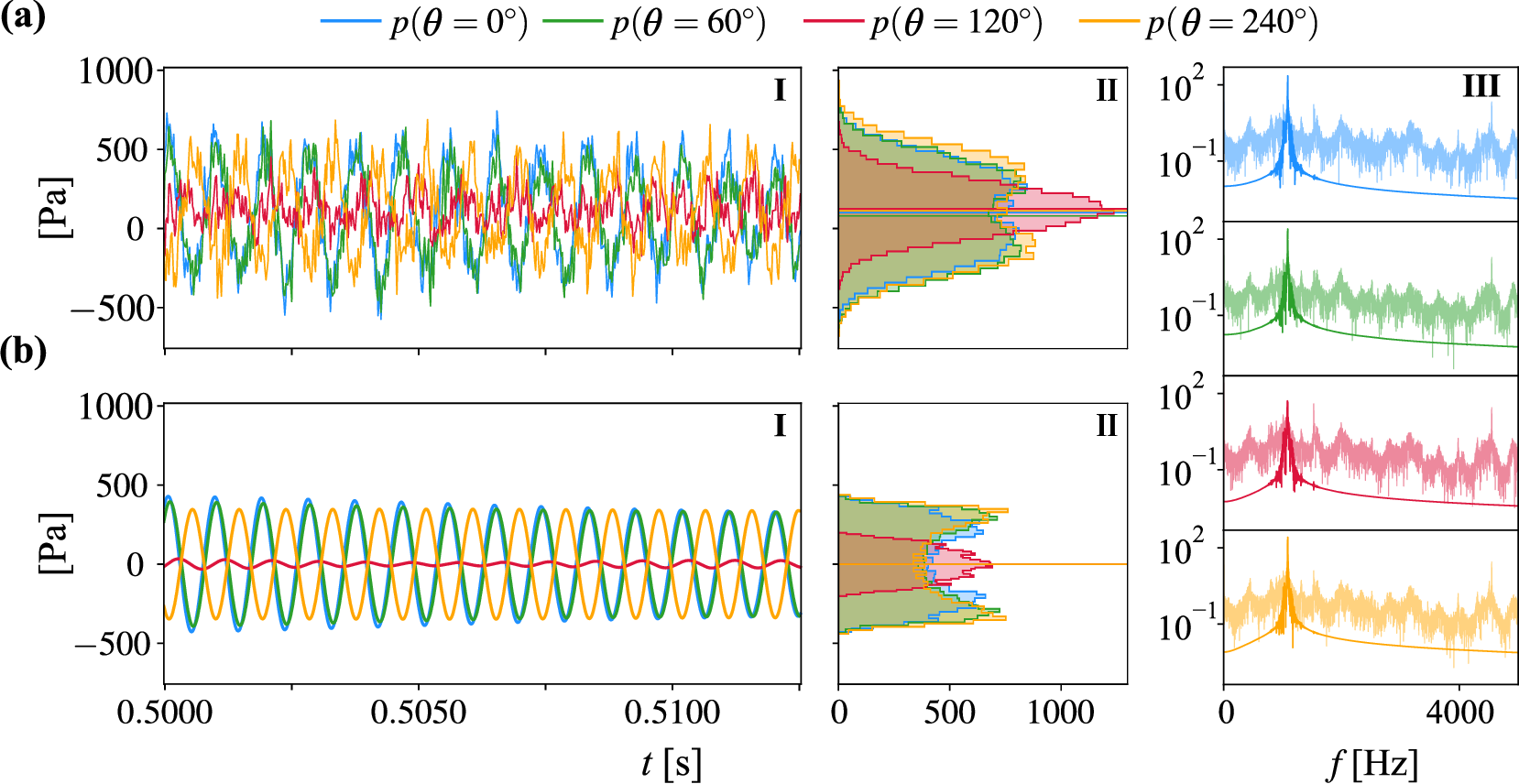}
    \caption{Pressure at the four measurement locations with equivalence ratio $\Phi=0.5125$ (thermoacoustically unstable configuration). (a) Raw experimental data, i.e., the observables, and (b) post-processed data, which is briefly referred to as the {\it presumed truth}. (I) Time series of the fast-varying oscillations, (II) histogram of the acoustic pressure in the time window for data assimilation, $t\in[0.5, 0.85]$, and {(III) comparison of the power spectral density of the raw data (light colour) and the presumed truth (dark colour).} The horizontal lines in (II) indicate the mean of the {histograms}.  The thermoacoustic limit cycle is a standing acoustic mode with a nodal line near $\theta=120^\circ$ (mono-modal histogram). }
    \label{fig:experimental_data}
\end{figure}
{
Figure~\ref{fig:experimental_data} shows the time series, histograms and power spectral density (PSD) of the raw  and post-processed experimental data at $\Phi=0.5125$. The PSD of the post-processed data is obtained after applying a Butterworth band-pass filter to the raw data around the frequency of instability, which approximately isolates the frequencies between 1050 and 1150 Hz (Figure~\ref{fig:experimental_data}{III})}.
The  pressure measurements have a mean value that is different from zero because of large-scale flow structures, which are not correlated to the thermoacoustic dynamics. The non-zero mean pressure is referred to as ``measurement shift'', as further explained in \S\ref{sec:uncertainties_definitions}. 
In a digital twin that is coupled with sensors' measurements in real time, the data are the  raw pressure signals that are assimilated into the low-order model (Figure~\ref{fig:experimental_data}{a}). Because the raw pressure contains aleatoric noise, measurement shift, and turbulent flow fluctuations, we refer to the post-processed data (Figure~\ref{fig:experimental_data}{b})  as the ``presumed ground truth'' or ``presumed acoustic state'', that is, the  physical acoustic state that is the presumed target prediction.

\subsection{A qualitative and physics-based low-order model}\label{sec:LOM_annular}
The dynamics of the azimuthal acoustics can be qualitatively modelled by a one-dimensional wave-like equation \citep[e.g.,][]{faure_noiray_symmetry_2020, indlekofer_noiray_spontaneous_2022}
\begin{align}\label{eq:annular_wave}
    &\dfrac{\partial^2p}{\partial t^2} + \zeta\dfrac{\partial p}{\partial t} - \left[1+\epsilon\cos{\left(2(\theta-\Theta_\epsilon)\right)}\right]\dfrac{c^2}{r^2}\dfrac{\partial^2p}{\partial \theta^2} = (\gamma - 1) \dfrac{\partial \dot{q}}{\partial t}, \\[1em] \label{eq:Qdot}
    &\text{with}\quad
    (\gamma - 1) \dfrac{\partial \dot{q}}{\partial t} = \beta\left[1+c_2\cos{(2(\theta-\Theta_\beta))}\right]p - \kappa p^3,
\end{align}
where 
$p$ is the acoustic pressure;
$\zeta$ is the acoustic damping;
$c$ is the speed of sound;
$r$ is the mean radius of the annulus;
$\gamma$ is the heat capacity ratio;
$\epsilon$, and $\Theta_\epsilon$ are the amplitude and phase of the reactive symmetry, respectively~\citep{indlekofer_noiray_spontaneous_2022}; 
and  
$\dot{q}$ is the coherent component of the fluctuations of the heat release rate~\citep{noiray_linear_2017}, which is divided into a nonlinear cubic term, which models the saturation of the flame response weighted by  the  parameter $\kappa$; and a linear response to the acoustic perturbations,  which is weighted  by the heat release strength $\beta$,  the resistive asymmetry intensity, $c_2$, and direction of maximum root mean square acoustic pressure, $\Theta_\beta$. 
The flame response model~\eqref{eq:Qdot} is an accurate approximation in the vicinity of a Hopf bifurcation~\citep[e.g.,][]{lieuwen_statistical_2003, noiray_linear_2017}. 
%
%
To further reduce the complexity of~\eqref{eq:Qdot}, 
we project the acoustic pressure on the degenerate pair of eigenmodes of the homogeneous wave equation~\citep{noiray_investigation_2011, noiray_dynamic_2013, magri_ARFM_2023}
\begin{equation}\label{eq:p_as_theta}
    p(t, \theta) = \eta_a(t) \cos\theta + \eta_b(t)\sin\theta,
\end{equation}
where $\eta_a$ and $\eta_b$ are the acoustic velocity amplitudes.  
Substituting~\eqref{eq:p_as_theta} into \eqref{eq:annular_wave} and averaging 
yield the governing equations, which consist of a set of nonlinearly coupled oscillators 
\begin{subequations}\label{eq:etas}
\begin{align}
    \ddot{\eta}_a = -&\omega^2\left[\eta_a \left(1 +  \frac{\epsilon}{2}\cos{(2 \Theta_\epsilon)}\right) + \eta_b \frac{\epsilon}{2} \sin{(2\Theta_\epsilon)}\right] +\\ \nonumber
    &\dot{\eta}_a\left[2\nu+\frac{c_2\beta}{2}\cos{(2\Theta_\beta)}-\frac{3\kappa}{4}\left(3\eta_a^2+\eta_b^2\right)\right]+
    \dot{\eta}_b\left[\frac{c_2\beta}{2}\sin{(2\Theta_\beta)}-\frac{3}{2}\kappa\eta_b\eta_a\right]\\
    \ddot{\eta}_b = -&\omega^2\left[\eta_b \left(1 -  \frac{\epsilon}{2}\cos{(2 \Theta_\epsilon)}\right) + \eta_a \frac{\epsilon}{2} \sin{(2\Theta_\epsilon)}\right] +\\  \nonumber
    &\dot{\eta}_b\left[2\nu-\frac{c_2\beta}{2}\cos{(2\Theta_\beta)}-\frac{3\kappa}{4}\left(3\eta_b^2+\eta_a^2\right)\right]+
    \dot{\eta}_a\left[\frac{c_2\beta}{2}\sin{(2\Theta_\beta)}-\frac{3}{2}\kappa\eta_b\eta_a\right]
\end{align}
\end{subequations}
where
$\nu=(\beta-\zeta)/2$ is the linear growth rate of the pressure amplitude in  absence of asymmetries, i.e., when $c_2=0$~\citep{indlekofer_noiray_spontaneous_2022}.  
Equation \eqref{eq:etas} can be written in a compact notation with a nonlinear state-space formulation 
\begin{align}
\label{eq:compact_etas}
\arraycolsep=1.4pt
\left\{
\begin{array}{rcl}
      \totald\vect{\phi}&=& \mathcal{F}\left(\vect{\phi, \alpha} \right)  \totald t \\[.5em]
     \vect{q} &=& \mathcal{M}(\vect{\theta}, \vect{\phi}),
\end{array}
\right.
\end{align}
where  
$\vect{\phi}=[\eta_a; \dot{\eta}_a; \eta_b; \dot{\eta}_b]\in \mathbb{R}^{N_\phi}$ is the state vector; 
$\mathcal{F}:\mathbb{R}^{N_\phi}\rightarrow\mathbb{R}^{N_\phi}$ is the nonlinear operator  that represents  \eqref{eq:etas}; 
and
$\vect{\alpha}=[\nu; c_2\beta; \omega; \kappa; \epsilon; \Theta_\beta; \Theta_\epsilon]\in \mathbb{R}^{N_\alpha}$ are the system's parameters (the operator $[~;~]$ indicates vertical concatenation, and we use column vectors throughout the manuscript). 
The vector $\vect{q}$ describes the model observables, which are the acoustic pressures  at the azimuthal locations~$\vect{\theta}= \{0^\circ, 60^\circ, 120^\circ, 240^\circ\}$. The model observables are computed through the measurement operator 
$\mathcal{M}:\mathbb{R}^{N_\phi}\rightarrow\mathbb{R}^{N_q}$, which maps the state variables into the observable space through \eqref{eq:p_as_theta}.

\section{Real-time data assimilation}\label{sec:data_assimilation}

Real-time data assimilation makes qualitatively models quantitatively accurate every time that sensors' measurements (data) become available~\citep{magri_doan_2019}. 
At a measurement time $t_k$, the model's state, $\vect{\phi}_k$, and parameters, $\vect{\alpha}_k$, are inferred by combining two sources of information, that is, the measurement, $\vect{d}_k$, and the model forecast, $\vect{q}_k$. 
A robust inference process must also filter out both epistemic and aleatoric uncertainties (\S\S\ref{sec:uncertainties_definitions},\ref{sec:4953489fn320}). The result of the assimilation is the \textit{analysis} state (\S\S\ref{sec:r-EnKF},\ref{sec:ensemble_framework}), which provides a statistically optimal estimate of the physical quantity that we wish to predict, which is the \textit{truth} $\vect{d}^{\dagger}_k$. As explained in \S\ref{sec:exp_setup}, we define the post-processed acoustic pressure as the presumed ground truth (Figure~\ref{fig:experimental_data}). 
We drop the time subscript $k$ unless it is necessary for clarity. 
\begin{figure}
    \centering
    \includegraphics[width=\linewidth]{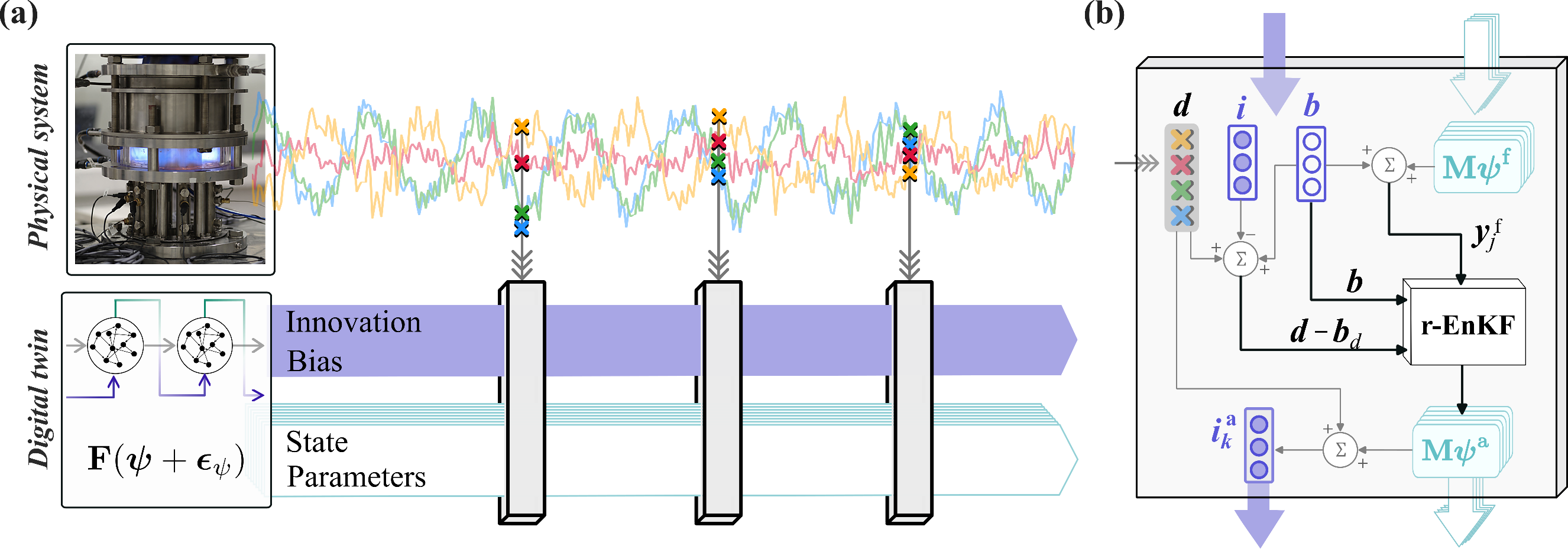}
    \caption{{Schematic of the proposed digital twin framework. (a) The physical and digital systems evolve in time (left to right) independently. The digital system is composed of a physical model with uncertain state and parameters, and an estimator of the  model bias and innovations (from which the measurement shift is estimated). When sensor data (crosses) become available, they are combined with the estimates from the digital twin using the regularized ensemble Kalman filter (r-EnKF) to update the digital system. (b) Diagram of the r-EnKF update  performed sequentially every time that data become available.}}
    \label{fig:dt_schematic}
\end{figure}

\subsection{Aleatoric and epistemic uncertainties}\label{sec:uncertainties_definitions}

First, we discuss the statistical hypotheses on the aleatoric uncertainties. 
The aleatoric uncertainties contaminate the state and parameters as 
\begin{equation}
    \vect{\phi}  + \vect{\epsilon}_\phi = \vect{\phi}^\mathrm{t}, \quad
    \vect{\alpha} + \vect{\epsilon}_\alpha = \vect{\alpha}^\mathrm{t},
\end{equation}
where $\mathrm{t}$ indicates the true quantity (which is unknown). 
The aleatoric uncertainties are modelled as Gaussian distributions   
\begin{equation}
    \vect{\epsilon}_\phi \sim \mathcal{N}(\vect{0}, \matr{C}_{\phi\phi} ), \quad \vect{\epsilon}_\alpha \sim \mathcal{N}(\vect{0}, \matr{C}_{\alpha\alpha} ),
\end{equation}
where $\mathcal{N}(\vect{0}, \matr{C})$ is a normal distribution with zero mean and covariance $\matr{C}$. 
Second, we discuss model biases, which are epistemic uncertainties~\citep{novoa_magri_inferring_2024}. The model bias may arise from modelling assumptions and  approximations in the operators $\mathcal{F}$ and $\mathcal{M}$. The \textit{true} bias of a model is defined as 
\begin{equation}\label{eq:bias_definition}
  {\vect{b}^\mathrm{t} = \vect{d}^\mathrm{t}} -\mathbb{E}(\vect{q}),
\end{equation}
which is the expected difference between the true observable and the model observable. 
(If the model is unbiased, ${\vect{d}^\mathrm{t}} =\mathbb{E}(\vect{q})$.) 
{
The true bias is unknown (colloquially, an ``unknown unknown''~\citep{novoa_magri_inferring_2024}) because we do not have access to the ground truth. Therefore, we need to estimate it from the information available from the data. We referred to as the \textit{presumed true} bias as $\vect{b}^\dagger  = \vect{d}^\mathrm{t}-\mathbb{E}(\vect{q})$, and to the bias estimated from a model as $\vect{b}$. 
With this, the model equations, which define the first source of information on the system, are
\begin{align}\label{eq:compact_etas_noise}
\arraycolsep=1.4pt
\left\{
\begin{array}{rcl}
     \totald\vect{\phi}&=& \mathcal{F}\left(\vect{\phi+\vect{\epsilon}_\phi, \alpha+\vect{\epsilon}_\alpha} \right) \totald t \\[.5em]
      \vect{y} &=& \mathcal{M}(\vect{\theta}, \vect{\phi}) + \vect{b}+ \vect{\epsilon}_q
\end{array}
\right.
\end{align}
where $\vect{y}$ is the bias-corrected model observable, and $\vect{\epsilon}_q \sim \mathcal{N}(\vect{0}, \matr{C}_{qq})$. 
The set of equations \eqref{eq:compact_etas_noise} is not closed until we define an estimator  for the model bias (\S\ref{sec:ESN}). 
}\\

The second source of information on the system is the data measured by the sensors. 
Experimental data are affected by both aleatoric noise and measurement shifts (\S\ref{sec:exp_setup}). 
{
The \textit{true} measurement shift is 
\begin{equation}\label{eq:bias_d_definition}
   \vect{b}_d^\mathrm{t} = \mathbb{E}(\vect{d}) - \vect{d}^\mathrm{t}. 
\end{equation}
Because the ground truth is not known, the best estimate on the model bias is the 
presumed true measurement shift $\vect{b}_d^\dagger  = \vect{d}^\dagger-\mathbb{E}(\vect{d})$ and the estimated measurement shift is $\vect{b}_d$. With this, we define the measurements at a time instant as
\begin{equation}\label{eq:observations}
   \vect{d} + \vect{b}_d + \vect{\epsilon}_d =  \vect{d}^\mathrm{t}. 
\end{equation}
}
In the problem under investigation, $\vect{d}$ is the raw acoustic data, $\vect{d}^{\dagger}$ is is the post-processed data, and $\vect{b}_d$ is the non-zero mean of the raw data (see~\S\ref{sec:exp_setup}). 
The aleatoric noise affects the measurement as $\vect{\epsilon}_d \sim \mathcal{N}(\vect{0}, \matr{C}_{dd})$. 
For simplicity, we assume that the measurement errors are statistically independent, i.e.,  $\matr{C}_{dd}$ is a diagonal matrix with identical diagonal entries $\sigma_d$.\\

Both the model bias and measurement shift are unknown {\it a priori}. To infer them, we analyse  the residuals between the forecast and the observations, which are also known as \textit{innovations}~\citep{dee_todling_dataGEOS_2000,haimberger_homogenization_2007}, which are defined as
\begin{equation}\label{eq:expected_i}
   \vect{i} = \vect{d} - \vect{q} \quad\Rightarrow\quad \mathbb{E}(\vect{i}) = \vect{b}_d + \vect{b}.
\end{equation}
Therefore, the expected innovation is the sum of the measurement and model biases as defined in \eqref{eq:bias_definition} and \eqref{eq:bias_d_definition}. 
The relation between the innovations $\vect{i}$ and the biases $\vect{b}$ and $\vect{b}_d$ will be essential for the design of the bias estimator in~\S\ref{sec:ESN}.

{\begin{table}
\centering
\caption{Summary of the terminology.}
\label{tab:terminology}
\begin{tabular}{@{}cl@{}}
\toprule
Symbol & Description \\ \midrule
$\vect{d}^\mathrm{t}$ & Ground truth (unknown). \\
$\vect{d}^\dagger$ & Presumed truth, obtained by applying a Butterworth filter to the raw data. \\
$\vect{d}$ & Raw experimental data from microphones~\eqref{eq:observations}. \\\midrule
$\vect{b}^\mathrm{t}, \vect{b}_d^\mathrm{t}$ & True model bias and measurement shift (unknown). Eqs.~\eqref{eq:bias_definition} and \eqref{eq:bias_d_definition}. \\
$\vect{b}^\dagger, \vect{b}_d^\dagger$ &  Presumed true model bias and measurement shift. \\
$\vect{b}^{\,}, \vect{b}_d$ &  Estimates of the model bias and measurement shift (\S~\ref{sec:ESN}).\\ 
\end{tabular}
\end{table}}

\subsection{Augmented state-space formulation}\label{sec:4953489fn320}

We define the augmented state vector  $\vect{\psi}=[\vect{\phi}; \vect{\alpha};\vect{q}]$, which comprises the state variables, the thermoacoustic parameters and  the model observables to  formally have a linear measurement operator $\matr{M}$, which simplifies the derivation of data assimilation methods~\citep[e.g.,][]{novoa_magri_2022}.
The augmented form of the model~\eqref{eq:compact_etas_noise} yields 
\begin{align}\label{eq:problem}
\arraycolsep=1.4pt
\left\{
\begin{array}{rcl}
     \totald\begin{bmatrix}
     \vect{\phi}\\
     \vect{\alpha}\\
     \vect{q}
 \end{bmatrix} &=& 
 \begin{bmatrix}
     \mathcal{F}(\vect{\phi}+\vect{\epsilon}_\phi,\vect{\alpha}+\vect{\epsilon}_\alpha)\\
     \vect{0}_{N_\alpha}\\
     \vect{0}_{N_q}
 \end{bmatrix} 
{\totald t}  \\[2em]
\vect{y} &=& \vect{q} + \vect{b}  + \vect{\epsilon}_{q}
\end{array}
\right.
\;\leftrightarrow  \;
\left\{
\begin{array}{rcl}
\totald\vect{\psi} &=& \matr{F}\left(\vect{\psi} +\vect{\epsilon}_\psi\right){\totald t}  \\[1em]
\vect{y} &=& \matr{M}\vect{\psi} + \vect{b}  + \vect{\epsilon}_{q}
\end{array}
\right. {, \;\forall\,t \neq {t_d}}
\end{align}
where 
{$ t_d$ are the times when the assimilation is performed, }
$\matr{F}(\vect{\psi})$ and $\vect{\epsilon}_\psi$ are the augmented nonlinear operator and aleatoric uncertainties, respectively; 
$\matr{M} = \left[\matr{0}~\big|~\mathbb{I}_{N_q}\right]$ is the linear measurement operator, which consists of the vertical concatenation of a matrix of zeros, $\matr{0}\in\mathbb{R}^{N_q\times (N_\phi+N_\alpha)}$, and the identity matrix, $\mathbb{I}_{N_q}\in\mathbb{R}^{N_q\times N_q}$; 
and $\vect{0}_{N_\alpha}$ and $\vect{0}_{N_q}$ are vectors of zeros (because the parameters are constant in time, and $\vect{q}$ is not integrated in time but it is only computed at the analysis step).

\subsection{Stochastic ensemble framework}\label{sec:ensemble_framework}

Under the Gaussian assumption, the inverse problem of finding the states, $\vect{\phi}$, and parameters, $\vect{\alpha}$, given some observations, $\vect{d}$, would be be solved by the Kalman filter equations if the operator $\mathcal{F}$ were linear~\citep{kalman_new_1960}. Thermoacoustic oscillations, however, have nonlinear dynamics (see \eqref{eq:annular_wave}). 
Stochastic ensemble methods are suitable for nonlinear systems because they do not require to propagate the covariance, in contrast to other sequential methods, e.g., the extended Kalman filter~\citep{evensen_book_2009,novoa_magri_2022}. 
Stochastic ensemble filters track in time $m$ realizations of the augmented state $\vect{\psi}_j$ to estimate the mean and covariance, respectively  
\begin{subequations}\label{eq:ensemble_stat}
\begin{align}
	 \mathbb{E}(\vect{\psi})&\approx\bar{\vect{\psi}}=\dfrac{1}{m}\sum^m_{j=1}{\vect{\psi}_j} 
	 \\
	 \matr{C}_{\psi\psi} = \begin{bmatrix}
				\matr{C}_{\phi\phi}  & \matr{C}_{\phi\alpha}& \matr{C}_{\phi q} \\
				\matr{C}_{\alpha \phi}  & \matr{C}_{\alpha\alpha}& \matr{C}_{\alpha q} \\
				\matr{C}_{q\phi}  & \matr{C}_{q\alpha}& \matr{C}_{qq} \\
			\end{bmatrix}
			&\approx\dfrac{1}{m-1}\sum^m_{j=1}(\vect{\psi}_i-\bar{\vect{\psi}})\otimes(\vect{\psi}_i-\bar{\vect{\psi}}),
\end{align}
\end{subequations}
where $\otimes$ is the dyadic product. 
%
%
Because the forecast operator $\mathcal{F}$ is nonlinear, the Gaussian prior may not remain Gaussian after the model forecast, and $\mathbb{E}(\mathcal{F}(\vect{\psi}))\neq\mathcal{F}(\bar{\vect{\psi}})$. However, the time between analyses $\Delta t_d$ is assumed small enough such that the Gaussian distribution is not significantly distorted~\citep{evensen_book_2009,yu_combined_2019}. \\

{Finally, we approximate the model bias as $\vect{b}\approx\vect{d}^\mathrm{t} - \matr{M}\bar{\vect{\psi}}$, thus, the sum of the biases is approximately equal to the mean of the innovations, i.e.,
\begin{equation}\label{eq:innovation_approx}
    \bar{\vect{i}} = \vect{d} - \matr{M}\bar{\vect{\psi}} \approx \vect{b}_d + \vect{b}.
\end{equation}
}


\subsection{The bias-regularized  ensemble Kalman filter} \label{sec:r-EnKF}

The objective function in a bias-regularized ensemble data assimilation framework contains three norms~\citep{novoa_magri_inferring_2024} 
\begin{align}\label{eq:renkf_cost_func}
\mathcal{J}(\vect{\psi}_j) = &\left\|\vect{\psi}_j-\vect{\psi}_j^\mathrm{f}\right\|^2_{\matr{C}^{\mathrm{f}^{-1}}_{\psi\psi}} +
 \left\|{\vect{y}}_j-\vect{d}_j\right\|^2_{\matr{C}^{-1}_\mathrm{dd}}+\gamma\left\|\vect{b}_j\right\|^2_{\matr{C}^{-1}_\mathrm{bb}}, \quad \mathrm{for} \quad j=0,\dots,m-1
\end{align}
where 
the superscript `f' indicates `forecast'; the operator $\left\|\cdot\right\|^2_{\matr{C}^{-1}}$ is the $L_2$-norm weighted by the semi-positive definite matrix ${\matr{C}^{-1}}$; $\gamma\geq0$ is a user-defined bias regularization factor; and $\vect{b}_j$ is the model bias of each ensemble member. For simplicity, we define the bias in the ensemble mean, i.e., $\vect{b}_j = \vect{b}$ for all $j$ (which we estimate with an echo state network in \S\ref{sec:ESN})~\citep{novoa_magri_inferring_2024}. 
From left to right, the norms on the right-hand-side of \eqref{eq:renkf_cost_func} measure 
(1) the spread of the ensemble prediction, 
(2)  the distance between the bias-corrected estimate and the observables, 
and 
(3) the model bias.  
The analytical solution of the bias-regularized ensemble Kalman filter (r-EnKF) in~\eqref{eq:r-EnKF}, which globally minimizes the cost function \eqref{eq:renkf_cost_func} with respect to $\vect{\psi}_j$, is 
\begin{subequations}\label{eq:r-EnKF}
\begin{align}
    \vect{\psi}_j^\mathrm{a} &= 
    \vect{\psi}_j^\mathrm{f} + 
    \matr{K}_r \left[\left(\mathbb{I}+ \matr{J}\right)\T\left(\vect{d}_j + \vect{b}_d - \vect{y}_j^\mathrm{f}\right) - \gamma \matr{C}_{dd}\matr{C}^{-1}_\mathrm{bb}\matr{J}\T\vect{b}\right], \quad j=0,\dots,m-1
\end{align}
with
\begin{align}
\matr{K}_r = \matr{C}_{\psi\psi}^\mathrm{f}\matr{M}\T\left[\matr{C}_{dd} + (\mathbb{I}+ \matr{J})\T(\mathbb{I}+ \matr{J})\matr{M}\matr{C}_{\psi\psi}^\mathrm{f}\matr{M}\T + \gamma \matr{C}_{dd}\matr{C}^{-1}_\mathrm{bb}\matr{J}\T\matr{J}\matr{M}\matr{C}_{\psi\psi}^\mathrm{f}\matr{M}\T\right]^{-1},
\end{align}
\end{subequations}
where  
 `a'  stands for `analysis', i.e, the optimal state of the assimilation; 
 we assimilate each ensemble member with a different  $\vect{d}_j\sim\mathcal{N}(\vect{d}, \matr{C}_{dd})$ to avoid covariance underestimation in ensemble filters~\citep{burgers_evensen_analysis_1998}; 
$\matr{K}_r$ is the regularized Kalman gain matrix; and 
$\matr{J}= \totald \vect{b}/\totald\matr{M}\vect{\psi}$ is the Jacobian of the bias estimator. 
Formulae~\ref{eq:r-EnKF}  generalize the analytical solutions of \citet{novoa_magri_inferring_2024} to the case of biased measurement. We prescribe  $\matr{C}_{dd} = \matr{C}_{bb}$ because the model  bias is defined in the observable space.  We use $\gamma$ to tune the norm of the bias in \eqref{eq:renkf_cost_func} \citep{novoa_magri_inferring_2024}. 
The optimal state and parameters are 
\begin{align}\label{eq:rEnKF_simple}
    \begin{bmatrix}
        \vect{\phi}_j^\mathrm{a}\\[1em]
        \vect{\alpha}_j^\mathrm{a}
    \end{bmatrix}
    = 
    \begin{bmatrix}
        \vect{\phi}_j^\mathrm{f}\\[1em]
        \vect{\alpha}_j^\mathrm{f}
    \end{bmatrix} + 
    \overbrace{
    \begin{bmatrix}
        \matr{C}_{\phi q}^\mathrm{f}\\[1em]
        \matr{C}_{\alpha q}^\mathrm{f}
    \end{bmatrix}
        \left\{\matr{C}_{dd}+(\mathbb{I}+ \matr{J})\T(\mathbb{I}+ \matr{J})\matr{C}_{qq}^\mathrm{f} +\gamma \matr{J}\T\matr{J}\matr{C}_{qq}^\mathrm{f}\right\}^{-1}
    }^{\text{Regularized Kalman gain,}\, \matr{K}_r}      \dots   \nonumber\\ 
    \dots\Big[\left(\mathbb{I}+ \matr{J}\right)\T\overbrace{\left(\vect{d}_j + \vect{b}_d - \vect{y}_j^\mathrm{f}\right)}^\text{Corrected innovation} - \gamma \matr{J}\T\vect{b}\Big]
\end{align}

The r-EnKF defines a `good' analysis from a biased model if  the unbiased state $\vect{y}$ matches the truth, and the model bias $\vect{b}$ is small relative to the truth. The underlying assumptions of this work are that (i) our low-order model is qualitatively accurate such that the model bias $\vect{b}$ has a small norm; and (ii) the sensors are properly calibrated.   
In the limiting case when the assimilation framework is unbiased (i.e., $\vect{b}=\vect{0}$, and $\vect{b}_d =\vect{0}$), the r-EnKF \eqref{eq:rEnKF_simple} becomes the bias-unregularized EnKF (\ref{app:classic_EnKF}).

\section{Reservoir computing for inferring model bias and measurement shift}\label{sec:ESN}

To apply the r-EnKF~\eqref{eq:rEnKF_simple} we must provide an estimate of the  model bias and the measurement shift. 
We employ an echo state network (ESN) for this task. ESNs are suitable for real-time data assimilation because  
(i) they are recurrent neural networks, i.e., they are designed to learn temporal dynamics in data; 
(ii) they are based on reservoir computing, hence they are universal approximators \citep{GRIGORYEVA2018495};
(iii) they are general nonlinear auto-regressive models \citep{aggarwal2018neural}; and
(iv)  training an ESN consists of solving a  linear regression problem, which provides a global minimum without back propagation.
In this work, we generalize the implementation of~\citet{novoa_magri_inferring_2024} to account for both model biases and measurement shift (\S\ref{sec:ESN_arch}). We propose a training dataset generation based on time-series correlation  (\S\ref{sec:ESN_train}).  

\subsection{Echo state network architecture and state-space formulation}\label{sec:ESN_arch}

We propose an ESN that simultaneously estimates $\vect{b}$ and $\vect{b}_d$ (where `f' indicates that the biases are forecast quantities, which are equivalent to the actual biases at convergence). 
Importantly, the biases $\vect{b}$ and $\vect{b}_d$ are not directly measurable in real-time, but we have access to the innovation, which is linked to the biases through~\eqref{eq:innovation_approx}. 
Once we have information on the innovation and the model bias (the measurement shift), we can estimate the measurement shift (the model bias)  through Eq.~\ref{eq:innovation_approx}.  \\

Figure~\ref{fig:ESN_partial_observations}{a} represents pictorially the proposed ESN at  time $t_k$, with its three main components: 
(i) the input data, which are  the mean analysis innovations, i.e, $\bar{\vect{i}}^\mathrm{a}_k=\vect{d}_k-\matr{M}\bar{\vect{\psi}}_k^\mathrm{a}$;  
(ii) the reservoir, which is a high-dimensional state is  characterized by the sparse reservoir matrix $\matr{W}\in\mathbb{R}^{N_{r}\times N_{r}}$ and  vector
$\vect{r}_k\in\mathbb{R}^{N_{r}}$ ($N_r\gg N_q$ is the number of neurons in the reservoir states); 
and 
(iii) the outputs, which are the mean innovation and the model bias at the next time step, i.e., $\bar{\vect{i}}_{k+1}$ and $\vect{b}_{k+1}$.  
The inputs to the network are a subset of the output (i.e., the innovation only). 
The sparse input matrix $\matr{W}_{\mathrm{in}}\in\mathbb{R}^{N_{r}\times (N_q+1)}$ maps the physical state into the reservoir, and the output matrix $\matr{W}_{\mathrm{out}}\in\mathbb{R}^{ N_q\times(N_r+1)}$ maps  the reservoir state back to the physical state. 
\begin{figure}
    \centering
    \includegraphics[width=\textwidth]{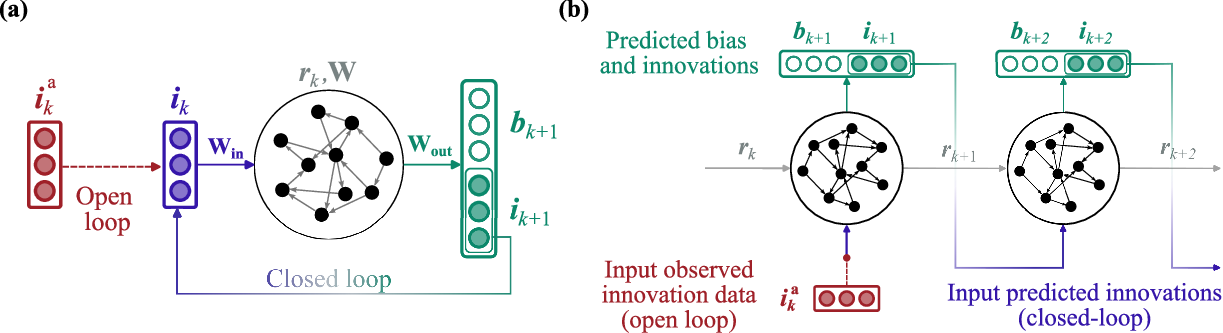}
    \caption{Schematic representation of the proposed echo state network architecture for model bias and innovation prediction. (a) Compact architecture showing the open-loop and closed-loop configurations; and (b) unfolded architecture starting at an analysis step is at $t_k$, in which there is one open-loop step followed by a closed-loop forecast.   
    In open loop the input to the ESN is the analysis mean innovation   $\bar{\vect{i}}^\mathrm{a}_{k}=\vect{d}_k-\matr{M}\bar{\vect{\psi}}_k^\mathrm{a}$, i.e., the difference between the raw acoustic pressure data and the analysis model estimate.  
    The outputs from the ESN are (i) the innovation  at the next time step $\bar{\vect{i}}_{k+1}$ (which becomes the input to the next time step in closed loop configuration); and (ii) the model bias   $\vect{b}_{k+1}$, i.e., an estimate of the difference between the presumed acoustic state $\vect{d}_{k+1}^\dagger$ and the model estimate $\vect{q}_{k+1}$.}
    \label{fig:ESN_partial_observations}
\end{figure}
Further, 
Figure~\ref{fig:ESN_partial_observations}{a}  shows the two forecast settings of the echo state network: open-loop, which is performed when data on the innovations are available, and the closed-loop, in which the ESN runs autonomously using the output as the input in the next time step.   
Figure~\ref{fig:ESN_partial_observations}{b} shows the unfolded architecture starting at   time $t_k$  when observations of the innovations are available (i.e., at the analysis step).  
At this time, the reservoir is re-initialized with the analysis innovations such that the input to the ESN is $\bar{\vect{i}}^\mathrm{a}_{k}=\vect{d}_k-\matr{M}\bar{\vect{\psi}}_k^\mathrm{a}$. From this, the ESN estimates at the next time step $t_{k+1}$ the model bias and the innovation $\bar{\vect{i}}_{k+1}$, which is used as initial condition in the subsequent forecast step.  Mathematically, the equations that forecast the model bias and mean innovation  in time are
\begin{align}\nonumber
\label{eq:ESN_OG_partial}
        [\vect{b}_{k+1}; \bar{\vect{i}}_{k+1}] &= \matr{W}_{\mathrm{out}}\left[\vect{r}_{k+1}; 1\right]\\
        \vect{r}_{k+1} &= \textrm{tanh}\left(\sigma_\mathrm{in}\matr{W}_{\mathrm{in}}\left[\bar{\vect{i}}^\star_k\odot\vect{g}; \delta_r\right]+
        \rho\matr{W}\vect{r}_{k}\right), 
\end{align}
where  $\bar{\vect{i}}^\star=\bar{\vect{i}}^\mathrm{a}$ in open loop and $\bar{\vect{i}}^\star = \bar{\vect{i}}$ in closed loop; 
the ${\tanh(\cdot)}$ operation is performed element-wise; 
the operator $\odot$ is the Hadamard product, i.e.,~component-wise multiplication;  
and 
$\vect{g}=[g_1;\dots; g_{2N_q}]$ is the input normalizing term with $g_q^{-1} = \max{(\vect{u}_q)}-\min{(\vect{u}_q)}$, i.e.,~$g_q^{-1}$ is the range of the $q^\mathrm{th}$ component of the training data $\matr{U}$; 
$\delta_r$ is a constant used for  breaking  the symmetry of the  ESN (we set $\delta_r=0.1$)~\citep{huhn_magri_learning_2020}.  
We define $\matr{W}_{\mathrm{in}}$ and $\matr{W}$ as sparse and randomly generated, with a connectivity of 3 neurons~\citep[see][for details]{jaeger2004harnessing}.  
We compute the matrix  $\matr{W}_{\mathrm{out}}$  during the training.

Lastly, the r-EnKF equations~\eqref{eq:r-EnKF} require the definition of the Jacobian of the bias estimator at the analysis step. Because at the analysis step the ESN inputs the analysis innovations with an open-loop step (see Figure~\ref{fig:ESN_partial_observations}{b}), the Jacobian of the bias estimator is equivalent to the negative Jacobian of the echo state network in open loop configuration (see~\ref{sec:Jacobian}).


\subsection{Training the network}\label{sec:ESN_train}

During training, the ESN is in open-loop configuration (Figure~\ref{fig:ESN_partial_observations}{a}).  The inputs to the reservoir are the training dataset $\matr{U}=\left[\vect{u}_0\,|\dots|\,\vect{u}_{N_\mathrm{tr}-1}\right]$, in which each time component $\vect{u}_k = [\vect{b}_\mathrm{u}(t_k); {\vect{i}}_\mathrm{u}(t_k)]$, the subscript `u' indicates  training data, and the operator $[~|~]$ indicates horizontal concatenation.  
Although we have information from the experimental data to train the network, the optimal parameters of the thermoacoustic system are unknown.  Thus, we do not know \textit{a priori} the model bias and measurement shift. 
Selecting an appropriate training dataset is key to obtaining a robust ESN, which can estimate the model bias and innovations. 
We create a set of $L$ guesses on the bias and innovations from a single realization of the experimental data, which means that the ESN is not trained with the `true' bias (which is an unknown quantity in real time). 
The training data is generated from the experimental data as detailed in \ref{app:ESN_train_data}. In summary, the procedure is as follows.
\begin{enumerate}
    \item{Take measurements for a training time window $t_\mathrm{tr}$ of acoustic pressure data $\matr{D}_\mathrm{u}$, and estimate $\matr{D}^{\dagger}_\mathrm{u}$ by applying a Butterworth filter to the raw data $\matr{D}$ (see \S\ref{sec:azimuthal_TA}). }
    \item{Generate $L$ model estimates $\matr{Q}_{\mathrm{u}, l}$ of the acoustic pressure from \eqref{eq:etas}. {Each $\matr{Q}_{\mathrm{u}, l}$ has a different set of parameters, which are uniformly randomly generated (the parameters' ranges are reported in \ref{app:params}).}}
    \item {Correlate in time each  $\matr{Q}_{\mathrm{u}, l}$ with $\matr{D}_\mathrm{u}$.}
    \item{Compute the model bias and innovations as
$
    \matr{B}_{\mathrm{u},l}= \matr{D}^{\dagger}_\mathrm{u} - \matr{Q}_{\mathrm{u},l}
$
and  
$
    \matr{I}_{\mathrm{u},l} =  \matr{D}_\mathrm{u} - \matr{Q}_{\mathrm{u},l}. 
$
}
\end{enumerate}
Finally, we apply data augmentation to improve the network's adaptivity in a real-time assimilation framework. The total number of training time series in the proposed training method is $2L$. 
Training the ESN consists of finding the elements in $\matr{W}_\mathrm{out}$, which minimize the distance between the outputs obtained from an open-loop forecast step of the training data (the features) to the training data at the following time step (the labels). 
This minimization is solved by ridge regression of the linear system~\citep{lukovsevivcius_practical_2012} 
\begin{equation}
\label{eq:RidgeReg_ens}
    \left(\sum_{l = 0}^{2L-1}\matr{R}^{\,}_l\matr{R}_l\T + \lambda\, \mathbb{I}_{N_r+1}\right)\matr{W}_{\mathrm{out}}\T = \sum_{l=0}^{2L-1}\matr{R}_l^{\,}\matr{U}_l\T, 
\end{equation}
where 
$\lambda$ is the Tikhonov regularization parameter; and 
$\matr{R} = \left[[\vect{r}_0; 1]\,|\dots|\,[\vect{r}_{N_\mathrm{tr}-1}; 1]\right]$, with $\vect{r}_0=\vect{0}$ and  $\vect{r}_k$ are obtained with \eqref{eq:ESN_OG_partial} using the innovations of the training set as inputs. 
The summations over $2L$ can be performed in parallel to minimize computational costs. 
Finally, we tune the hyperparameters of the echo state network, i.e., the spectral radius $\rho$, the input scaling $\sigma_\mathrm{in}$ and the Tikhonov regularization parameter $\lambda$. We use a recycle validation strategy with Bayesian optimization for the hyperparameter selection~\citep[see][]{racca_magri_robust_2021}.

\section{Implementation}
{To summarize, four stages are necessary to design the real-time digital twin: 
\begin{description}
    \item[1. Initialization: ]{The ensemble $\vect{\psi}_j\; \mathrm{for}\; j = 1,...,m$, and the ESN are initialized using the parameters reported in Appendix \ref{app:params}.}
    \item[2. Forecast: ]{Time-march in parallel each ensemble member \eqref{eq:problem}, i.e., the system of thermoacoustic equations and parameters; and the ESN according to \eqref{eq:ESN_OG_partial} until observation data become available.}
    \item[3. Analysis: ]{Apply the bias-aware r-EnKF~\eqref{eq:r-EnKF}, which obtains the optimal combination between the unbiased model estimate and the observation data.  }
    \item[4. Re-initialization: ]{Update the state and parameters of the ensemble with the  analyses (i.e., $\vect{\psi}^\mathrm{a}$ and $\vect{\alpha}^\mathrm{a}$), and the ESN  with the analysis innovation (i.e., $\bar{\vect{i}}^\mathrm{a} = \vect{d} - \matr{M}\bar{\vect{\psi}}^\mathrm{a}$).  }
\end{description}
Steps (\textbf{2})-(\textbf{4}) are repeated sequentially as data become available. Once the assimilation process has ended, we forecast further the ensemble and the ESN to analyse the extrapolation and generalization capability on the estimated state, parameters and biases. 
}

\section{Performance of the real-time digital twin}\label{sec:results}
{The real-time digital twin has four components:
(i) data from sensors (\S\ref{sec:exp_setup})
(ii) the physics-based low order  model (\S\ref{sec:LOM_annular}), 
(iii) the echo state network to infer the model bias and measurement shift (\S\ref{sec:ESN}),  and 
(iv) the real-time and bias-regularized data assimilation method (\S\ref{sec:r-EnKF}) to twin sensors' data  with the low-order model to infer the state and model parameters. } 
We investigate the capability of the proposed real-time digital twin to predict autonomously the dynamics of azimuthal thermoacoustic oscillations. 
We analyse the digital twin's performance with respect to established methods: the bias-unregularized ensemble Kalman filter (EnKF)~\citep{novoa_magri_2022}, which is suitable for real-time assimilation; and Langevin-based regression~\citep{indlekofer_noiray_spontaneous_2022}, which is suitable for offline assimilation. 
The hyperparameters used to train the ESN, and the assimilation parameters are reported in \ref{app:params}.

\subsection{Time-accurate prediction}\label{sec:exp_results_time}

We focus our analysis on the equivalence ratio $\Phi=0.5125$, which is a thermoacoustically unstable system. 
Figure~\ref{fig:times0.5125}  shows the time evolution of the acoustic pressure at $\theta=60^\circ$.  
{
Prior to the assimilation, we train the ESN with training data over $t_\mathrm{tr}=0.167$ seconds. 
We start the assimilation at $0.5s$ to avoid using training data during the assimilation, and to have the ensemble in statistically stationary regime. In parallel, we initialize the ESN before the assimilation begins with 10 data points (i.e., the washout). }
After this, observations from the raw experimental data (red dots) are assimilated every $\Delta t_d=5.86\E{-4}$ seconds, which corresponds to approximately $0.64$ data points per acoustic period. {(The assimilation frequency was selected by down-sampling the raw data, which is recorded at 51.2~kHz, such that $\Delta t_d$ fulfils to the Shannon-Nyquist criterion.)} 
At a time instant, the measurement (the data point, red dot) is assimilated into the model, which adaptively updates itself through the r-EnKF and bias estimators (\S\S\ref{sec:r-EnKF},\ref{sec:ESN}). 
After that, the state and model parameters have been  updated, the low-order model~\eqref{eq:compact_etas} evolves autonomously until the next data point becomes available. 
This process mimics a stream of data coming from sensors on the fly. 
We assimilate 600 measurements during the assimilation window {of $0.35s$}, the model runs autonomously without seeing more data ($t>0.85$~s).

First, we analyse the performance of the bias-unregularized filter (EnKF) on state estimation (Figure~\ref{fig:times0.5125}{a}) with the corresponding parameter inference shown in Figure~\ref{fig:params0.5125} (dashed lines). 
\begin{figure}
    \centering
    \includegraphics[width=\textwidth]{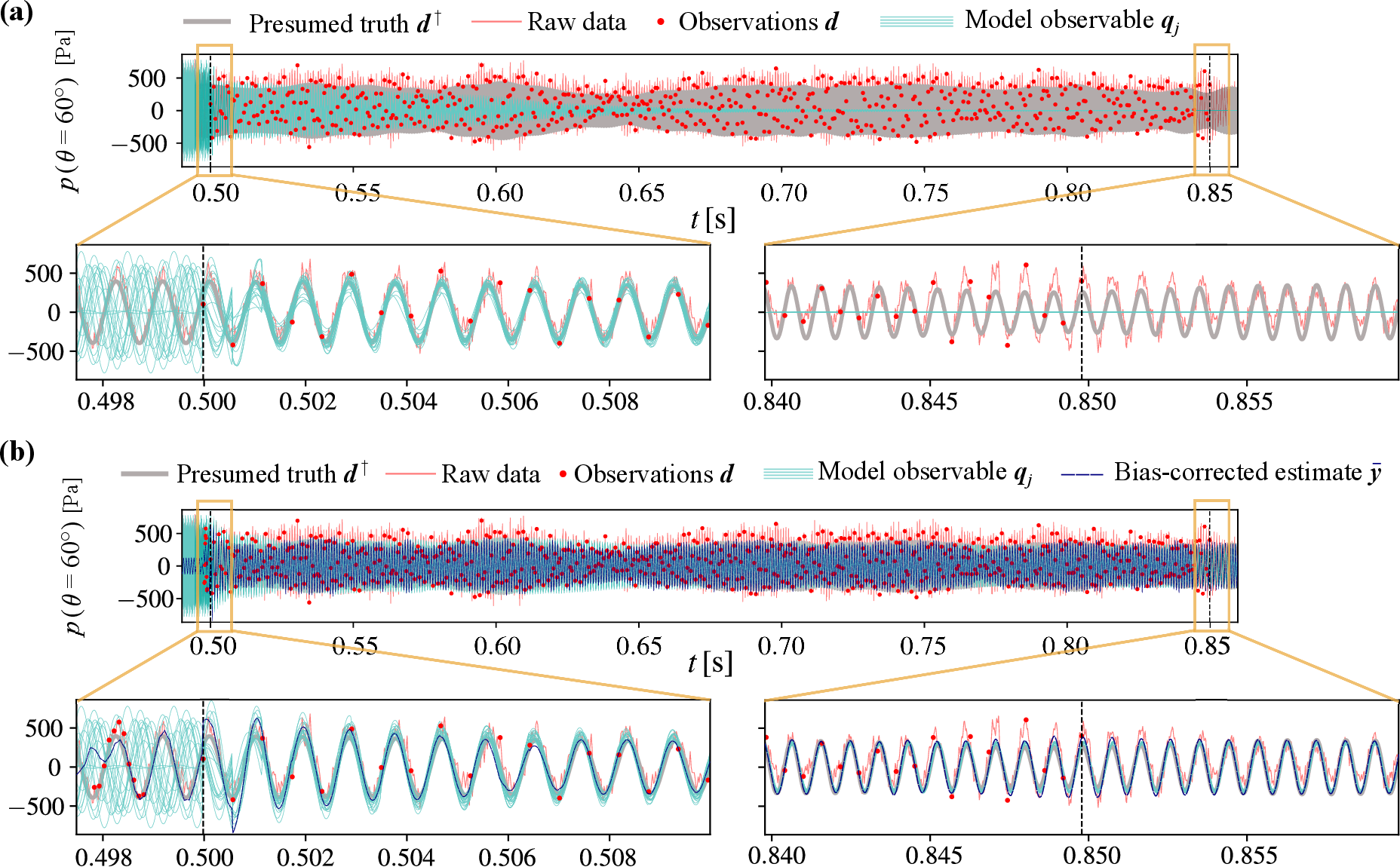}
    \caption{Real-time digital twin at $\Phi=0.5125$. 
    Time evolution of the thermoacoustic pressure (in Pa) at $\theta=60^\circ$  using (a) the bias-unregularized EnKF and (b) the proposed bias-regularized EnKF (r-EnKF). 
    Comparison between the presumed ground truth (thick grey line), the raw data (thick red line), the prediction from the ensemble filter (cyan lines), and in (b) the bias-corrected mean estimate (navy dashed line). The close ups show the start and end of the assimilation window, which is indicated by the vertical dashed lines. The observations are show in red circles. }
    \label{fig:times0.5125}
\end{figure}
\begin{figure}
    \centering
    \includegraphics[width=\textwidth]{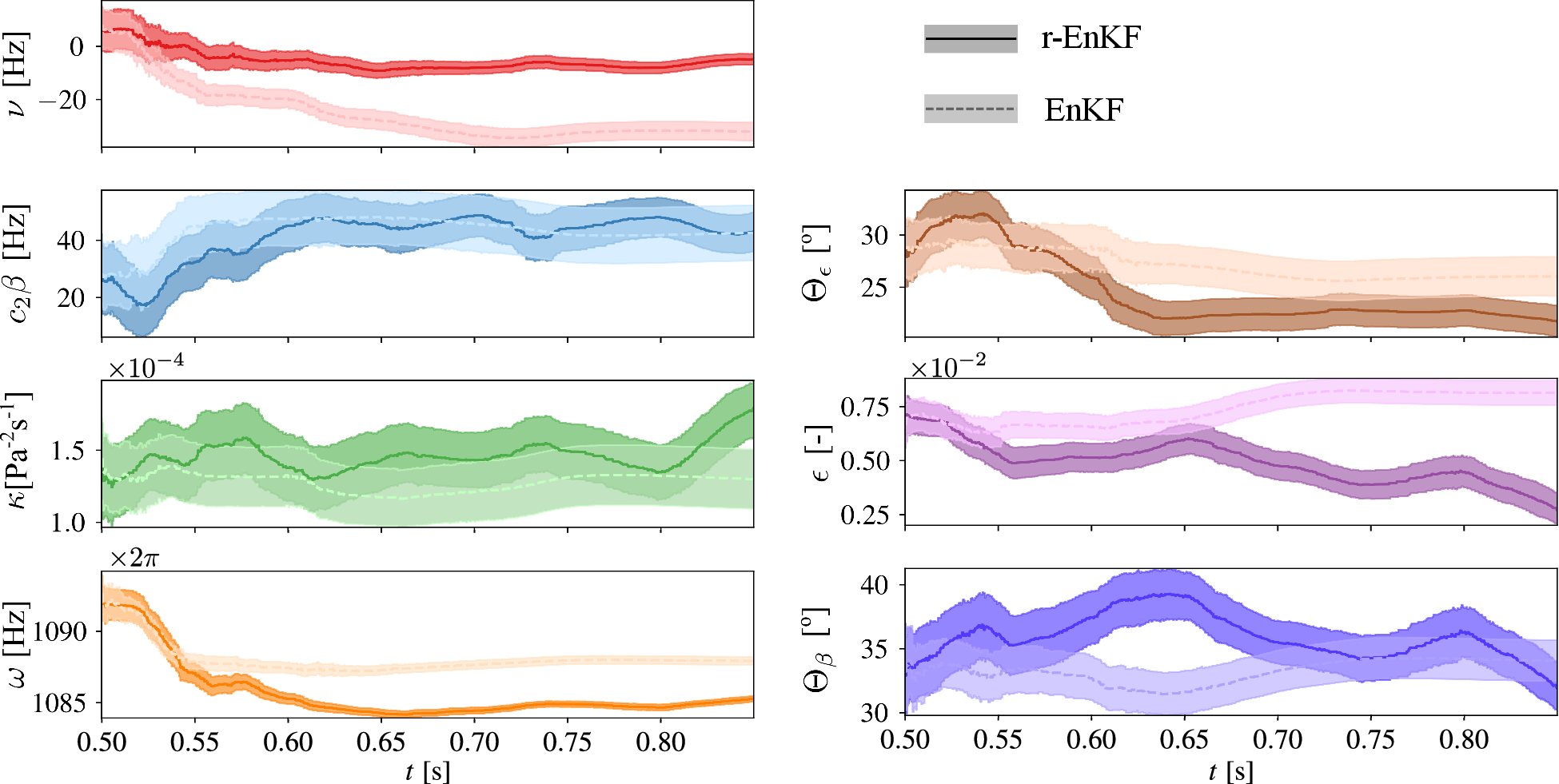}
    \caption{Thermoacoustic parameters with the bias-unregularized EnKF (dashed) and the bias-regularized  r-EnKF (solid). The lines and shaded areas indicate the ensemble mean and standard deviation, respectively. $\Phi=0.5125$.}
    \label{fig:params0.5125}
\end{figure} 
Although the system has a self-sustained oscillation, the bias-unregularized filter converges towards an incorrect solution, that is, a fixed point. 
On the other hand, the bias-regularized filter (r-EnKF) successfully learns the thermoacoustic model. 
Second, the r-EnKF filters out aleatoric noise and turbulent fluctuations. These fluctuations are the difference between the amplitudes of the raw data, which are the input to the digital twin, and the post-processed data, which are the hidden state that we wish to uncover  and predict (the presumed acoustic state), of up to approximately 200~Pa (close-ups at the end of assimilation in Figure~\ref{fig:times0.5125}). 
Third, after assimilation, the r-EnKF predicts the thermoacoustic limit cycle beyond the assimilation window. This means that the digital twin has learnt an accurate model of the system, which generalizes beyond the assimilation. The ensemble model observables $\vect{q}_j$ and bias-corrected ensemble mean $\bar{\vect{y}} = \bar{\vect{q}} + \vect{b}$ are almost identical at convergence in Figure~\ref{fig:times0.5125}{b}, which means that the magnitude of the model bias is small (approximately 20~Pa in amplitude,  see Figure~\ref{fig:PDF_0.5125}). 
Fourth, the r-EnKF infers the optimal system's thermoacoustic parameters, which are seven in this model (Figure~\ref{fig:params0.5125}). {The ensemble parameters at $t=0$ in  Figure~\ref{fig:params0.5125} are initialized on physical ranges from  the physical parameters used in the literature, which are computed by Langevin regression in~\citep{indlekofer_noiray_spontaneous_2022}.} From inspection of the optimal parameters, we can draw physical conclusions: 
(i) the parameters of the literature are not the optimal parameters of the model; 
(ii) the linear growth rate, $\nu$, angular frequency, $\omega$, the heat release strength weighted by the symmetry intensity, $c_2\beta$, and the phase of the reactive symmetry, $\Theta_\epsilon$, do not significantly vary at regime, which means that these parameters are constants and do not depend on the state; and 
(iii) the nonlinear parameter, $\kappa$, the amplitude of the reactive symmetry, $\epsilon$, and the direction of asymmetry, $\Theta_\beta$ have temporal variations that follow the modulation of the pressure signal envelope (Figure~\ref{fig:params0.5125}). 
Physically, this means that the optimal deterministic system that represents azimuthal instabilities is a time-varying parameter system~\citep{durbin_koopman_timeseries_2012}. By inferring time-varying parameters, we derive a deterministic system that does not require stochastic modelling.
We further analyse the thermoacoustic parameters for all the equivalence ratios in Figures~\ref{fig:nu_beta_ERs} and~\ref{fig:rest_of_params}.

To summarize, the presence of model bias and measurement shift makes the established bias-unregularized ensemble Kalman filter fail to  uncover and predict  the physical state from noisy data. In contrast, the bias-regularized filter (r-EnKF) infers the state, parameters, model bias, and measurement shift, which enable a time-accurate and physical prediction beyond the assimilation window.

\subsection{Statistics and biases}

We analyse the uncertainty and the statistics of the time series (\S\ref{sec:exp_results_time}) generated by the real-time digital twin. 
To do so, we define the normalized root-mean square (RMS) error of two time series $\vect{w}, \vect{z}$ with $N_q$ dimensions and $N_t$ time steps  as 
\begin{align}\label{eq:RMS}
    \mathrm{RMS}(\vect{w}, \vect{z}) &= \sqrt{\dfrac{\sum_{q}\sum_{k} \left({w}_q(t_k)- {z}_q(t_k)\right)^2}{\sum_{q} \sum_{k}\left({w}_q(t_k)\right)^2}}, \quad \mathrm{for} \;
    \begin{array}{l}
    q=0,\dots, N_q-1, \\[.5em]
    k=0,\dots,N_t-1. 
    \end{array}
\end{align}
\begin{figure}
    \centering
    \includegraphics[width=\textwidth]{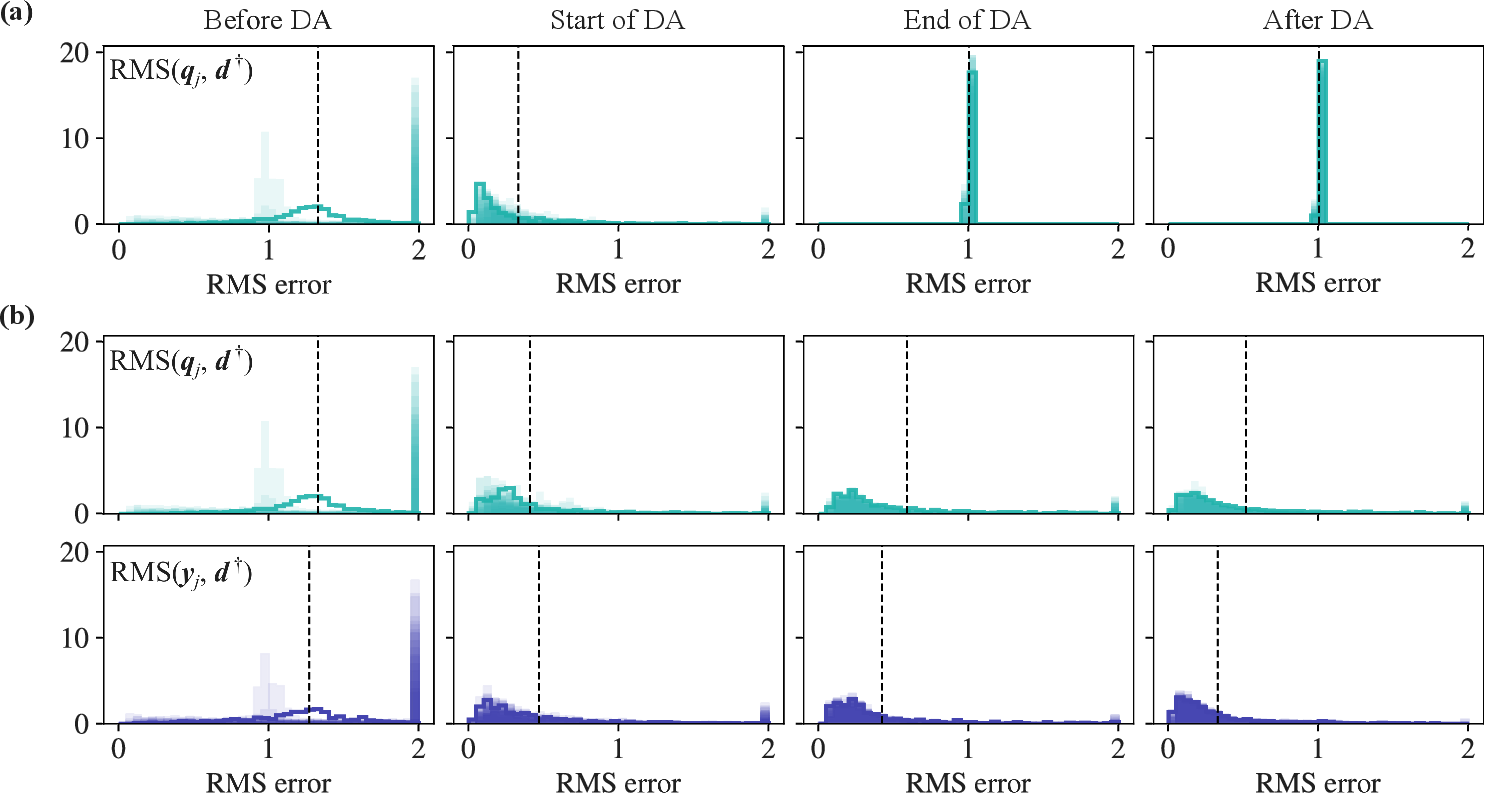}
    \caption{Normalized RMS errors between the presumed ground truth and the model prediction (cyan) and between the presumed ground truth and the bias-corrected prediction (navy) with (a) the bias-unregularized EnKF and (b) the r-EnKF. 
    The filled histograms show the RMS of each ensemble member, the thick lines show the RMS of the ensemble mean, and the vertical dashed line indicates the mean of the RSM  error.    
    The errors are computed at four stages of the assimilation:  $t\in[0.49, 0.50]$ (before DA),   $t\in[0.50, 0.51]$ (start of DA), $t\in[0.84, 0.85]$ (end of DA), and  $t\in[0.85, 0.86]$ (after DA). $\Phi=0.5125$.}
    \label{fig:RMS_0.5125}
\end{figure}
\begin{figure}
    \centering
    \includegraphics[width=\textwidth]{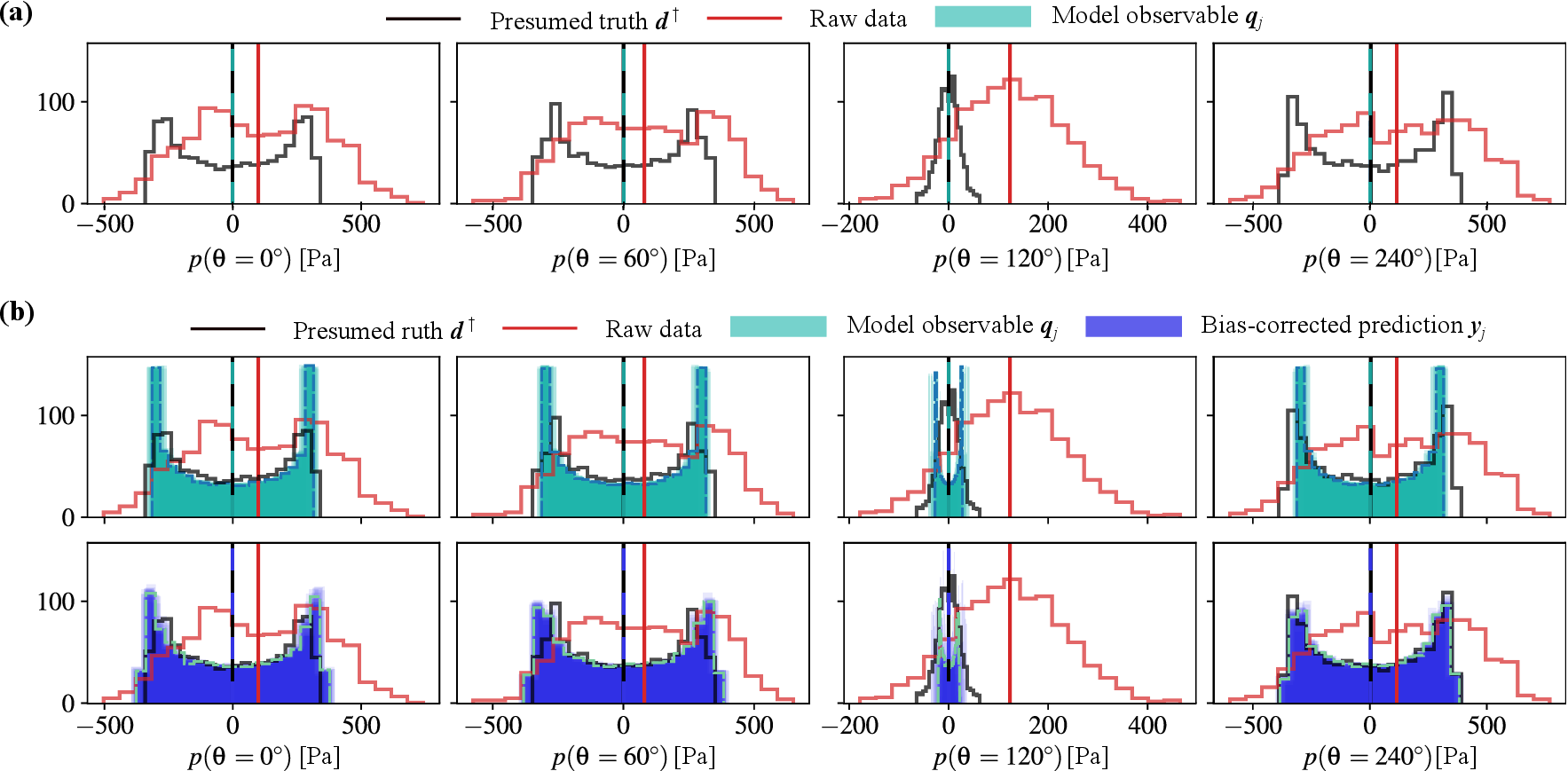}
    \caption{Histograms of the acoustic pressure after assimilation with (a) the bias-unregularized EnKF and (b) the r-EnKF, at the four observed azimuthal locations. Comparison between the presumed ground truth (black), the observations from raw data (red), the ensemble prediction (filled cyan) and its mean (dashed teal), and bias-corrected ensemble predictions (filled navy) and its mean (dashed light blue). The vertical lines indicate the mean of each of the distributions. $\Phi=0.5125$.} 
    \label{fig:PDF_0.5125}
\end{figure}
Figure~\ref{fig:RMS_0.5125} shows the RMS errors at four critical time instants in the assimilation process: 
before the data is assimilated (first column), which corresponds to the  initialization of the digital twin;
at the start and end of the data assimilation window (second and third columns, respectively); and 
after the data assimilation (fourth column), i.e., when the digital twin evolves autonomously to predict unseen dynamics (generalization). 
The inference of the model bias and measurement shift is key to obtaining a small generalization error. 
Although the bias-unregularized EnKF converges to an unphysical solution with a large RMS, 
the bias-regularized filter converges to a physical solution with a small RMS. 
As the assimilation progresses, the echo state network improves the prediction of the model bias because the RMS of the bias-corrected prediction (Figure~\ref{fig:RMS_0.5125}{b}, cyan) is smaller than the model estimate (Figure~\ref{fig:RMS_0.5125}{b}, navy). 
This is further evidenced by analysing the histograms of the time series for the four azimuthal locations  (Figure~\ref{fig:PDF_0.5125}).
The digital twin (bias-corrected solution, navy histograms) converges to the expected distribution  of the acoustic pressure (presumed truth)  (black histograms) despite the assimilated data is substantially contaminated by noise and turbulent fluctuations (red histograms). 
The bias-corrected histograms have a  zero mean, which means that the echo state network in the digital twin has correctly inferred both the model bias and the measurement shift. 
Both biases are shown in Figure~\ref{fig:biases}. 
The measurement shift, whose \textit{presumed true} value is known {\it a priori} $\vect{b}_d^\dagger = \langle \vect{d}-\vect{d}^\dagger\rangle$ (where the brackets $\langle~\rangle$ indicate time average), is exactly inferred by the ESN.
\begin{figure}
    \centering
    \includegraphics[width=\textwidth]{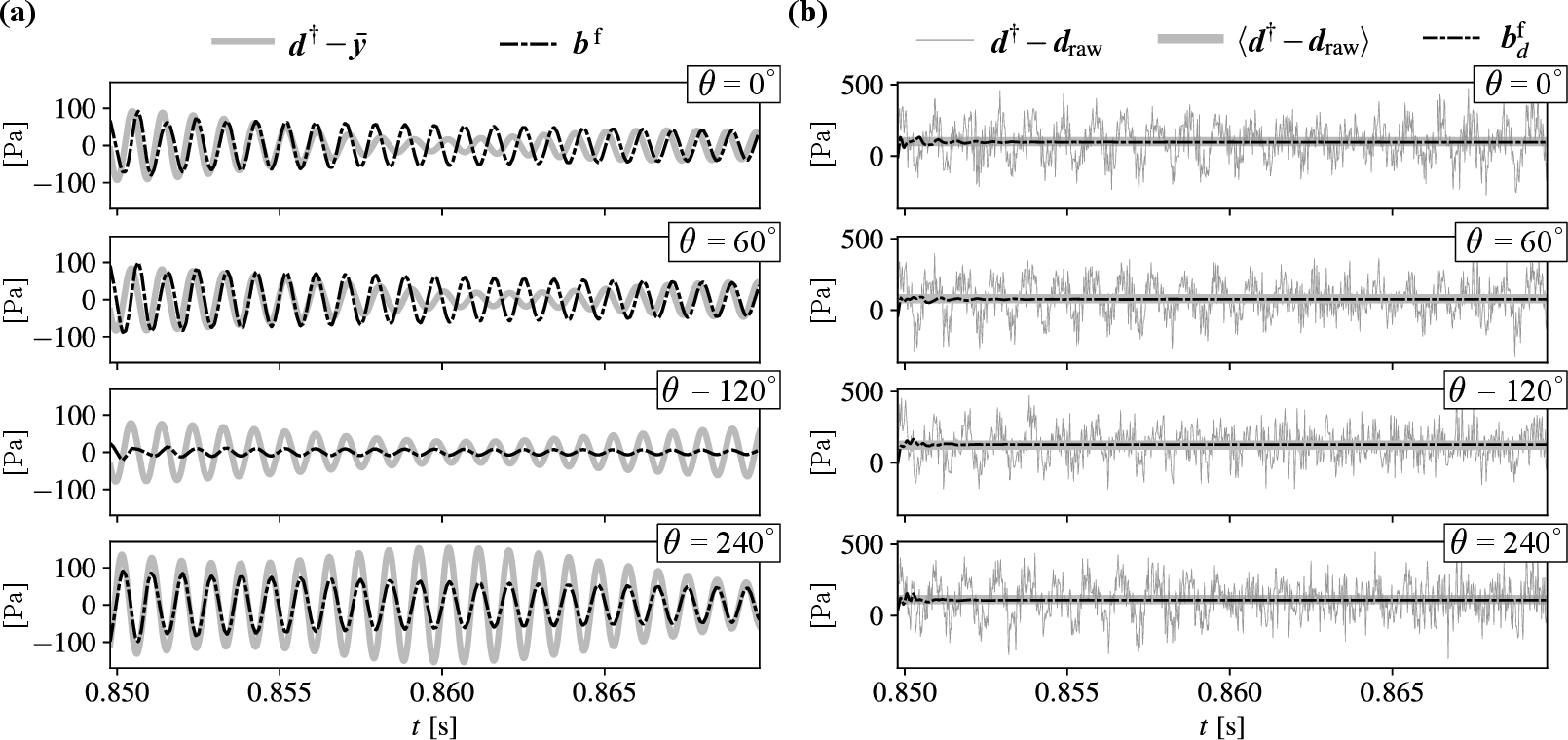}
    \caption{Bias-regularized r-EnKF. Comparison between the ESN inference of the biases and the  
    (a) actual acoustic state (dashed black) and the presumed model bias from data (thick grey). 
    (b) Measurement shift estimate (dashed black), difference between the raw and post-processed data (thin grey), and its time average, $\langle\cdot\rangle$ (thick grey). $\Phi=0.5125$.}
    \label{fig:biases}
\end{figure}
On the other hand, the {\it presumed true} value of the model bias is not known {\it a priori}, but it is presumed as $\vect{b} = \vect{d}^\dagger - \vect{q}^\dagger$, where $\vect{q}^\dagger$ is the presumed true model estimate, which is not known \textit{a priori}.  
The digital twin improves the knowledge that we have on the model bias by correcting  the presumed model bias. 

To summarize, the r-EnKF generalizes to unseen scenarios and extrapolates correctly in time. Key to the robust performance is the inference of the biases in the model and measurements.  
\subsection{Physical parameters and comparison with the literature}\label{sec:results-quat}

We analyse the system's physical parameters (\S\ref{sec:LOM_annular} and equation~\eqref{eq:etas})  for all the available equivalence ratios. 
We train an ESN for each equivalence ratio. 
The data assimilation parameters (assimilation window, frequency, ensemble size, etc.) are the same in all tests (\ref{app:params}).
We compare the low-order model parameters 
(i) inferred by the bias-unregularized EnKF,
(ii) inferred by the bias-regularized r-EnKF, and 
(iii) obtained with the state-of-the-art Langevin regression by~\citet{indlekofer_noiray_spontaneous_2022}. 
The linear growth rate, $\nu$, and the heat source strength weighted by the symmetry intensity, $c_2\beta$ are shown in Figure~\ref{fig:nu_beta_ERs}; and  the remaining parameters are shown in Figure~\ref{fig:rest_of_params}. 
The specific values parameters are listed in \ref{app:inferred_params}. 
 \begin{figure}
     \centering
     \includegraphics[width=.75\textwidth]{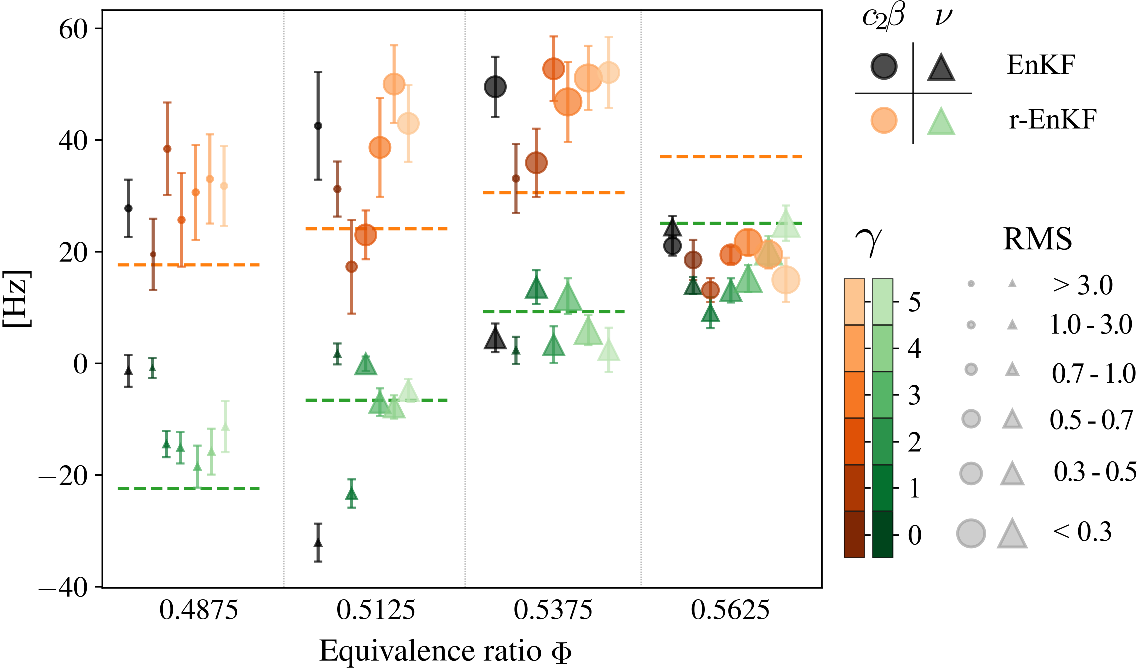}
     \caption{Comparison of the estimated values of the two stability parameters $c_2\beta$ (circles) and $\nu$ (triangles)  after data assimilation, obtained with the bias-unregularized EnKF (black) and the r-EnKF with different bias regularization parameters (colour maps). Lighter colours indicate stronger regularization; larger marker sizes indicate more accurate solutions, i.e., smaller RMS errors; and the error bars represent the ensemble standard deviation. 
     The dashed lines corresponds to the  parameters identified  by offline Langevin regression~\citep{indlekofer_noiray_spontaneous_2022}. 
     }
     \label{fig:nu_beta_ERs}
 \end{figure}
 \begin{figure}
     \centering
     \includegraphics[width=.9\textwidth]{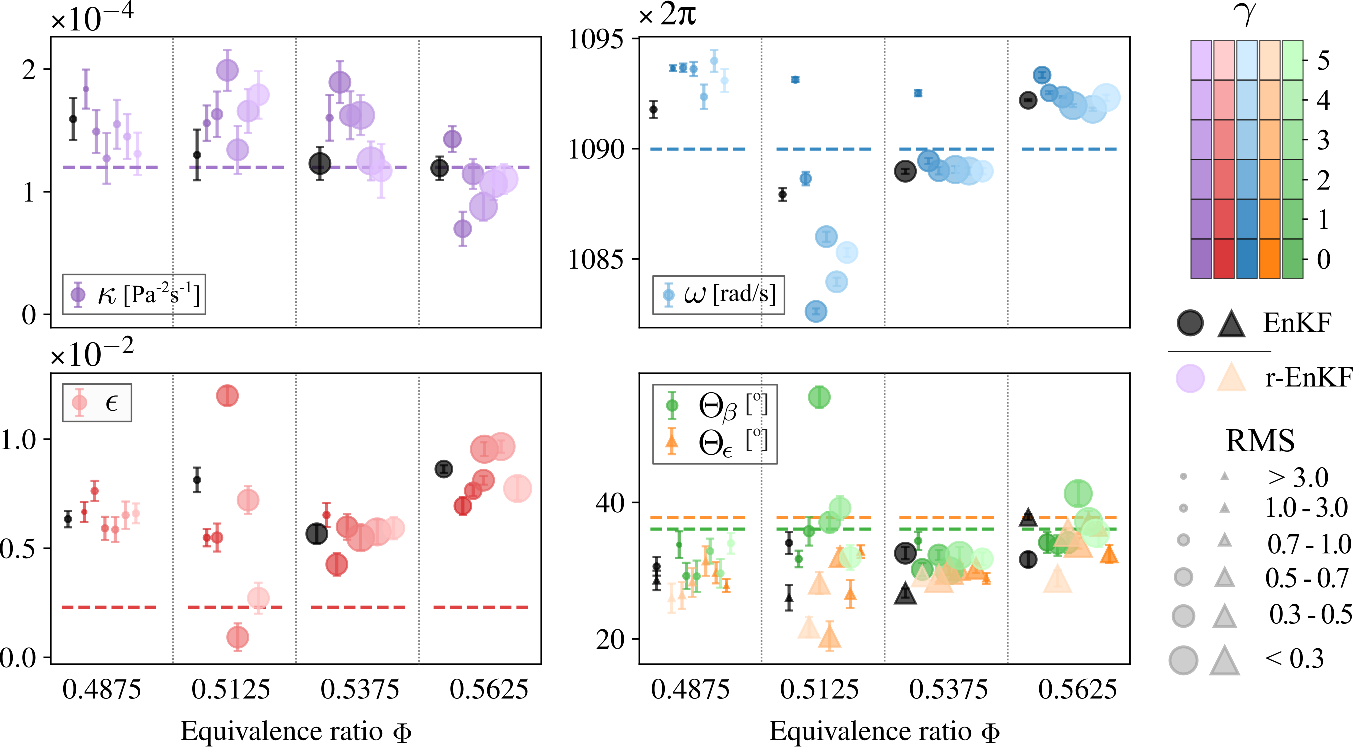}
     \caption{Same as Figure~\ref{fig:nu_beta_ERs} for the remaining physical parameters.}
     \label{fig:rest_of_params}
 \end{figure}

First, the bias-regularized EnKF  infers physical parameters with a small generalization error (the RMS errors are small, depicted with large markers). 
Large values of the bias regularization, i.e., $\gamma \gtrsim1$, provide the most accurate set of parameters.
This physically means that the low-order model is a qualitatively accurate model because the bias has a small norm (which corresponds to a large $\gamma$).
The bias-unregularized EnKF is outperformed in all cases. 
(The case with $\Phi=0.4875$ is a stable configuration with a trivial fixed point, which is why the parameters have a larger RMS.)
Second, we analyse the uncertainty of the parameters, which is shown by an error bar.
The height of the error bar is the ensemble standard deviation. 
The linear growth rate, $\nu$, and the angular frequency, $\omega$, are inferred with small uncertainties.
Physically, this means that the model predictions are markedly sensitive to small changes in the linear growth rate and angular frequency.
The prediction's sensitivity to the parameters change with the equivalence ratio. 
Physically, the large sensitivity of the linear growth rate (i.e., the system's linear stability) to the  operating condition is a characteristic feature of thermoacoustic instabilities, in particular in annular combustors, in which eigenvalues are degenerate~\citep[e.g.,][]{magri_ARFM_2023}.  
Third, there exists multiple combination of parameters that provide a similar accuracy.
For example, in the case $\Phi=0.5125$, there are different combinations of $\Theta_\beta, \epsilon, c_2\beta$, which yield a small RMS.  
Fourth, we compare the digital twin with the 
offline Langevin regression of~\citet{indlekofer_noiray_spontaneous_2022} (horizontal dashed lines in Figures~\ref{fig:nu_beta_ERs}, \ref{fig:rest_of_params}). 
The inferred parameters are physically meaningful because they lie within value ranges that are similar to the parameters of the literature. 
The digital twin simultaneously infers all the parameters in real time, which overcomes the limitations of Langevin regression, which needs to be performed on one parameter at a time and offline. 
{The parameters presented in Figures~\ref{fig:nu_beta_ERs} and~\ref{fig:rest_of_params} depend linearly on  $\Phi$. These linear dependencies are approximations of the ground truth and were inferred from the identified set of parameters at each $\Phi$ (see Figure 9 in \citet{indlekofer_noiray_spontaneous_2022}). When  tuned for a specific $\Phi$, the offline parameters lead to state statistics that more closely match the observables.}
The digital twin parameters are optimal because they are the minimizers of~\eqref{eq:renkf_cost_func}.  
%
%
The long-term statistics of the case $\Phi=0.5625$, which lives on a generalized Bloch sphere~\citep{magri_ARFM_2023}, is further analysed with the quaternion ansätz of~\citet{ghirardo_quaternion_2018}.
Figure~\ref{fig:quant+reind22e39j32} shows the acoustic state for $\Phi=0.5625$ from the raw data (panel b), 
and the proposed digital twin (panel c). 
{The long-term statistics are obtained by forecasting the model (without assimilation) using the identified parameters  (reported in \ref{app:inferred_params}).} 
The real-time digital twin infers a set of physical parameters, which correctly capture the physical state of the azimuthal acoustic mode. The nature angle has a bimodal distribution with $|\chi|<\pi/4$, which means that the thermoacoustic instability is a mixed mode. 
\begin{figure}
    \centering
    \includegraphics[width=\textwidth]{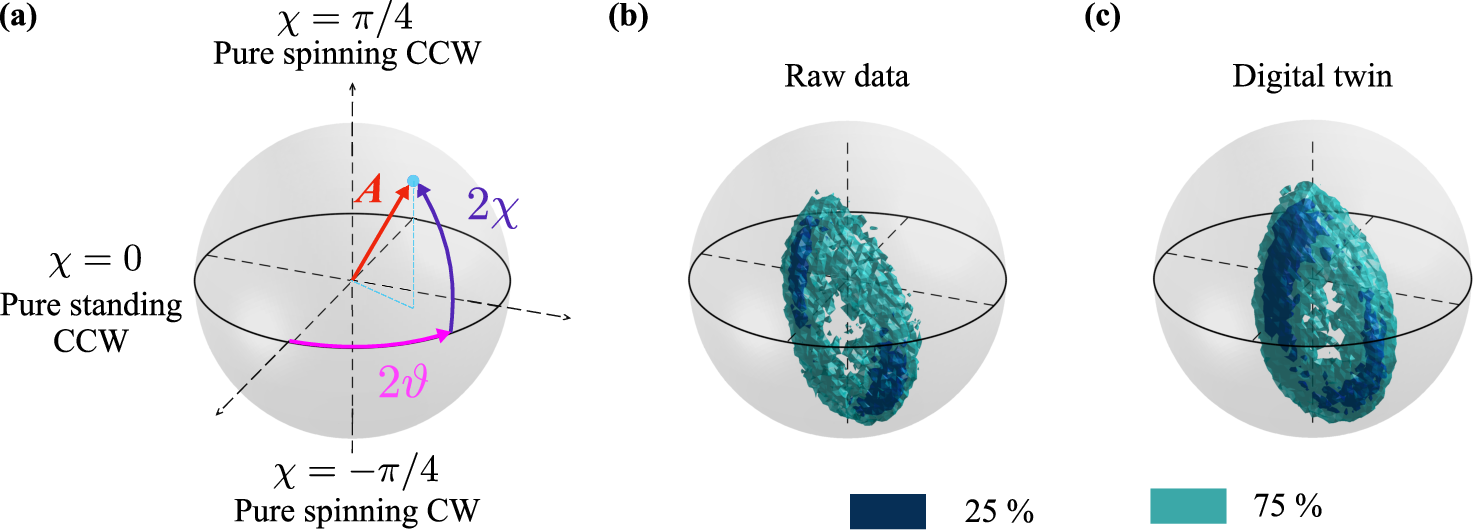 }
    \caption{Acoustic state for $\Phi=0.5625$. 
    (a) Bloch sphere for quaternion representation of the acoustic pressure $p(\theta, t) = A \cos{(\vartheta - \theta)} \cos{\chi} \cos{(\omega t + \varphi)} + A \sin{(\vartheta - \theta)} \sin{\chi} \sin{(\omega t + \varphi)}$, 
where 
$A$ is the slow-varying amplitude, i.e., the envelope of the acoustic pressure; 
$\varphi$, is the slow temporal phase drift, 
$\vartheta$ is  the position of the anti-nodal line, and 
$\chi$ is the nature angle~\citep{ghirardo_quaternion_2018, magri_ARFM_2023}. 
    (b) State from the raw data. 
    (c) State predicted by the proposed real-time digital twin.
    }
    \label{fig:quant+reind22e39j32}
\end{figure}

\subsection{{Generalizability}}

{
In this section, we propose a method for generalizing the digital twin to unseen experimental data, e.g., a different equivalence ratio.  In sections \S~\ref{sec:exp_results_time}-\ref{sec:results-quat}, we train a different ESN for each equivalence ratio. Here, we test the real-time digital twin  framework using an unified ESN, i.e., we train the ESN with data for $\Phi={0.4875, 0.5125, 0.5375}$ but not for the $\Phi_\mathrm{test}=0.5625$ case. The training data generation is identical to that detailed in \ref{app:ESN_train_data}, but we combine the training data for $\Phi={0.4875, 0.5125, 0.5375}$ to train one ESN, which has the same characteristics as those used in the previous analyses (\ref{app:params}). 
We increased the ensemble size to $m=40$ to account for the larger uncertainty and variability in the dynamics. 
Figure~\ref{fig:generalization1} shows the inferred linear growth rate $\nu$ and the heat source strength $c_2\beta$ for all $\Phi$. 
Notably, the digital twin for the equivalence ratio $\Phi_\mathrm{test}=0.5625$, which was unseen by the ESN, converges to similar parameters  to those in Figure~\ref{fig:nu_beta_ERs}.  We further analyse the test case  $\Phi_\mathrm{test}=0.5625$ in Figure~\ref{fig:generalization2}, which shows the histograms of the acoustic pressure after assimilation. The ESN infers the correct measurement shift to give a zero-mean prediction of the pressure, and a model bias which reduces the distance between the ensemble state estimate and the presumed truth. 
Overall, we conclude that the unified ESN successfully generalizes to data with unseen dynamics.
\begin{figure}
    \centering
    \includegraphics[width=.75\linewidth]{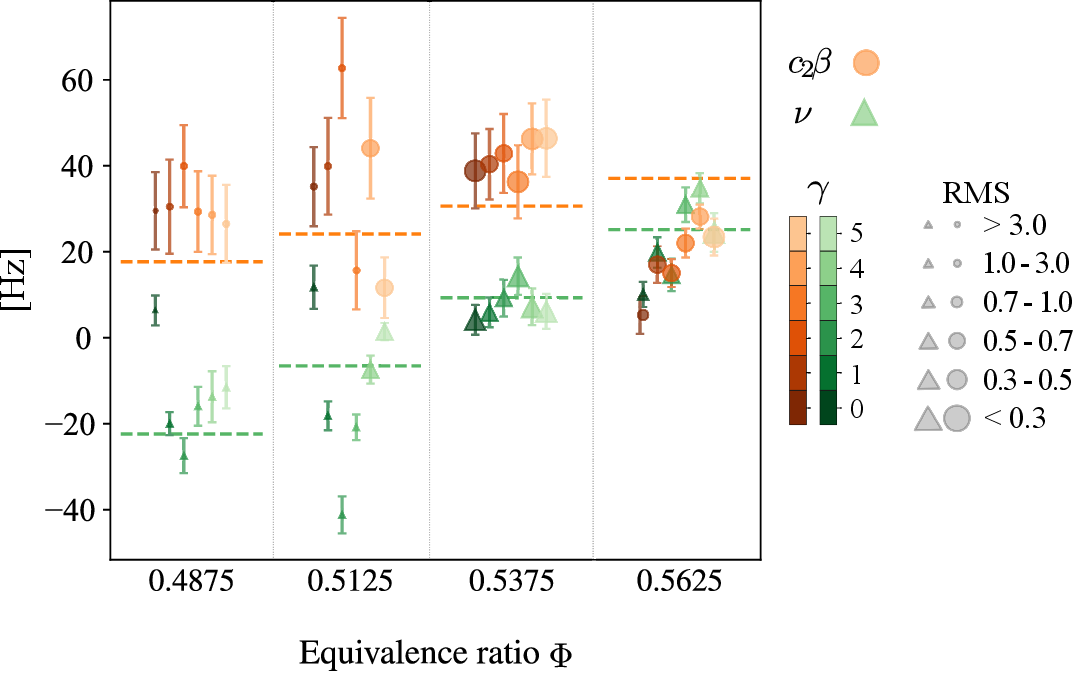}
    \caption{{Generalizability study. Estimated values of the two stability parameters $c_2\beta$ (circles) and $\nu$ (triangles)  after data assimilation with the r-EnKF using the same ESN, which is not trained with data for $\Phi=0.5625$. 
    Lighter colours indicate stronger bias regularization parameter; larger marker sizes indicate more accurate solutions, i.e., smaller RMS errors; and the error bars represent the ensemble standard deviation. 
     The dashed lines corresponds to the  parameters identified  by offline Langevin regression~\citep{indlekofer_noiray_spontaneous_2022}. }}
    \label{fig:generalization1}
\end{figure}
\begin{figure}
    \centering
    \includegraphics[width=\linewidth]{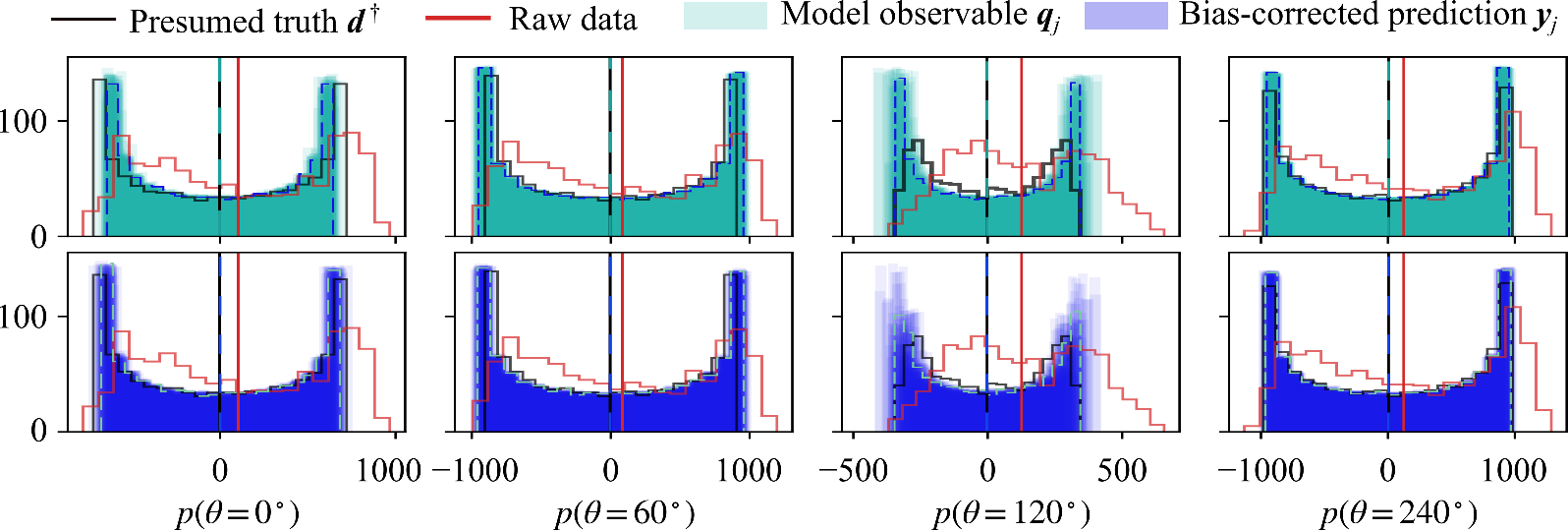}
    \caption{{Generalizability study. Histograms of the acoustic pressure at $\Phi=0.5625$ after assimilation with the r-EnKF, at the four observed azimuthal locations. The ESN used in the digital twin is not trained with data for $\Phi=0.5625$.
    Comparison between the presumed ground truth (black), the observations from raw data (red), the ensemble prediction (filled cyan) and its mean (dashed teal), and bias-corrected ensemble predictions (filled navy) and its mean (dashed light blue). The vertical lines indicate the mean of each of the distributions.} }
    \label{fig:generalization2}
\end{figure}
}

\section{Conclusions}\label{sec:conclusions}
We develop a real-time digital twin of azimuthal thermoacoustic oscillations of a hydrogen-based combustor for uncovering the physical state from raw data, and predicting the nonlinear dynamics in time.   
We identify the four ingredients requited to design a robust real-time digital twin: 
\begin{enumerate}[(i)]
\item  In a real-time context, we need to work with raw data as they come from sensors, therefore, we do not pre-process the data. The data we utilize is raw (the data contain both environmental and instrumental noise, and turbulent fluctuations) and sparse (the data are collected by four microphones).  
\item  A low-order model, which is deterministic, qualitatively accurate, computationally cheap, and physical.
\item  An estimator of both the model bias, which originates from the modelling assumptions and approximation of a low-order model; and measurement shift, which originates from  microphones recording the total pressure (which has a non-zero mean) instead of the acoustic pressure (which has zero mean). 
\item  A statistical data assimilation method that optimally combines the two sources of information on the system (i-ii) to improve the prediction on the physical states by updating the physical parameters every time that sensors’ data become available (on the fly, or real-time). 
\end{enumerate}

We propose a real-time data assimilation framework to infer the acoustic pressure, physical parameters, model biases and measurements biases simultaneously, which is the bias-regularized ensemble Kalman filter (r-EnKF). We find an analytical solution that minimizes the bias-regularized  data assimilation cost function \eqref{eq:renkf_cost_func}. We propose a reservoir computer (echo state network) and a training strategy to infer both the model bias a measurement shift, without making assumptions on their functional forms {\it a priori}. The real-time digital twin is applied to a laboratory hydrogen-based annular combustor for a variety of equivalence ratios. The data are treated as if they came from sensors on the fly, i.e., the pressure measurements are assimilated at a time step and then disregarded until the next pressure measurement becomes available. 
First, after assimilation of raw data (at 1707~Hz, i.e., approximately 0.6 data points per acoustic period), the model learns the correct physical state and optimal parameters. 
The real-time digital twin autonomously predicts azimuthal dynamics beyond the assimilation window, i.e., without seeing more data. This is in stark contrast with state-of-the-art methods based on the bias-unregularized methods, which do not perform well in the presence of model bias and measurement shift. 
Second, the digital twin uncovers the physical acoustic pressure from the raw data. Because the physical mechanisms are constrained in the low-order model, which originate from conservation laws, the digital twin acts as a physics-based filter, which removes aleatoric noise and turbulent fluctuations. 
Third, physically we find that azimuthal oscillations are governed by a time-varying parameter system, which generalizes existing models that have constant parameters, and capture only slow-varying variables.
Fourth, we find that the key parameters that influence the dynamics are the linear growth rates, and the angular frequency. Further, the digital twin generalizes to equivalence ratios at which current modelling approaches are not accurate.  
{
Fifth, to generalize the digital twin framework, we train an ESN with data from three of the four available equivalence ratios, and test the assimilation performance in the unseen scenario.  This unified ESN successfully estimates the model bias and measurement shift in unseen thermoacoustic dynamics with the regularized bias-aware data assimilation framework.   
}
This work opens new opportunities for real-time digital twinning of multi-physics problems. Current work is focused on exploring the role of sparsity in the data. 

\section*{Acknowledgments}
A.N. is supported partly by the EPSRC-DTP, UK, the Cambridge Commonwealth, European and International Trust, UK under a Cambridge European Scholarship, and Rolls Royce, UK. 
A.N. and L.M. acknowledge support from the UKRI AI for Net Zero grant EP/Y005619/1. 
L.M. acknowledges support from the ERC Starting Grant No. PhyCo 949388 and EU-PNRR YoungResearcher TWIN ERC-PI\_0000005.

\section*{Declaration of interest}
The authors report no conflict of interest.

\bibliographystyle{plainnat}
\bibliography{jfm-instructions}

\appendix

\section{The ensemble Kalman filter}\label{app:classic_EnKF}

The classical (i.e., bias-unregularized) ensemble Kalman filter (EnKF) updates each ensemble member as \citet{evensen_ensemble_2003}
\begin{align}\label{eq:EnKF}
    \vect{\psi}_j^\mathrm{a} 
    &= \vect{\psi}_j^\mathrm{f}+\matr{K}\left[\vect{d}_j - \matr{M}\vect{\psi}_j^\mathrm{f}\right], \quad j=0,\dots,m-1,\\ \nonumber
    &\mathrm{with} \quad \matr{K} = \matr{C}_{\psi\psi}^\mathrm{f}\matr{M}\T\left(\matr{C}_{dd}+\matr{M}\matr{C}_{\psi\psi}^\mathrm{f}\matr{M}\T\right)^{-1},
\end{align}
where $\matr{K}$ is the Kalman gain matrix, and 
the superscripts `f' and `a' indicate `forecast' and `analysis', respectively. 
By evaluating the measurement operator products, we can write the EnKF update for the states and parameters of each ensemble member as
\begin{align}\label{eq:EnKF_simple}
    \begin{bmatrix}
        \vect{\phi}_j^\mathrm{a}\\[1em]
        \vect{\alpha}_j^\mathrm{a}
    \end{bmatrix}
    &= 
    \begin{bmatrix}
        \vect{\phi}_j^\mathrm{f}\\[1em]
        \vect{\alpha}_j^\mathrm{f}
    \end{bmatrix} + 
    \underbrace{\begin{bmatrix}
        \matr{C}_{\phi q}^\mathrm{f}\\[1em]
        \matr{C}_{\alpha q}^\mathrm{f}
    \end{bmatrix}
    \left(\matr{C}_{dd}+\matr{C}_{qq}^\mathrm{f}\right)^{-1}}_\mathrm{Kalman \, gain, \, \matr{K}}\overbrace{\left(\vect{d}_j - \vect{q}_j^\mathrm{f}\right)}^\mathrm{Innovation}
\end{align}
As mentioned in \S\ref{sec:intro}, the EnKF is a bias-unregularized method. The estate and parameter update does not consider either the model bias or the measurement shift.

\section{Jacobian of the bias estimator}\label{sec:Jacobian}
The Jacobian of the bias estimator is equivalent to the negative Jacobian of the ESN in open loop configuration, such that 
\begin{equation}\label{eq:J_1}
    \matr{J} =\dfrac{\totald\vect{b}_{k+1}}{\totald\matr{M}\bar{\vect{\psi}}_k}= \dfrac{\totald\vect{b}_{k+1}}{\totald\bar{\vect{i}}_{k}^\mathrm{a}}\dfrac{\totald\bar{\vect{i}}_{k}^\mathrm{a}}{\totald\matr{M}\bar{\vect{\psi}}_k} =-\dfrac{\totald\vect{b}_{k+1}}{\totald\bar{\vect{i}}_k}= -\matr{W}_\mathrm{out}^{(1)}\left[\matr{T}\odot\left(\sigma_\mathrm{in}\matr{W}_\mathrm{in}^{(1)}\odot\matr{G}\right)\right], 
\end{equation}
where $\matr{G}=\left[\vect{g}~|\dots|~\vect{g}\right]\T\in\mathbb{R}^{N_q\times N_r}$; 
$\vect{1}\in\mathbb{R}^{N_r}$ is a vector of ones; 
and  $\matr{T}=\left[\vect{T}~|\dots|~\vect{T}\right]\in\mathbb{R}^{N_r\times N_q}$ with
\begin{align}\nonumber
\vect{T}=\vect{1}-\tanh^2\left(\sigma_\mathrm{in}\matr{W}_\mathrm{in}^{(1)}\left(\vect{b}_k\odot\vect{g}\right)+\sigma_\mathrm{in}\delta_r\matr{W}_\mathrm{in}^{(2)} +\rho\matr{W}\vect{r}_k\right). 
\end{align}
For details on the derivation, the reader is referred to~\citet{novoa_magri_inferring_2024}.

\section{Training the network}\label{app:ESN_train_data}

During training, the ESN is in open-loop configuration (Figure~\ref{fig:ESN_partial_observations}{a}).  The inputs to the reservoir are the training dataset $\matr{U}=\left[\vect{u}_0\,|\dots|\,\vect{u}_{N_\mathrm{tr}-1}\right]$, where each time component $\vect{u}_k = [\vect{b}_\mathrm{u}(t_k); {\vect{i}}_\mathrm{u}(t_k)]$ (the subscript `u' indicates  training data).  
Although we have information from the experimental data to train the network, the optimal parameters of the thermoacoustic system are unknown.  Thus, we do not know a priori the model bias and measurement shift. 
Selecting an appropriate training dataset is essential to obtain a robust echo state network which can estimate the model bias and innovations. 
We create a set of $L$ guesses on the bias and innovations from a single realization of the experimental data. This means that the ESN is not trained with the `true' bias.

The training data generation is summarized in Algorithm~\ref{alg:train_dataset}, and the procedure is as follows.  
First, we take measurements for a training time window $t_\mathrm{tr}$ of acoustic pressure data $\matr{D}_\mathrm{u}$, and we estimate $\matr{D}^{\dagger}_\mathrm{u}$ by applying a band-pass filter to the raw data $\matr{D}$ (see \S\ref{sec:azimuthal_TA}). 
Second, we draw $L$ sets of thermoacoustic parameters from uniform random distribution with lower and upper bounds $\vect{\alpha}^0(1-\sigma_L)$ and $\vect{\alpha}^0(1+\sigma_L)$, which are selected from an educated physical initial guess on the thermoacoustic model parameters {based on previous works~\citep{indlekofer_noiray_effect_2021, faure_noiray_imperfect_2021}. The ranges of the uniform distributions are reported in \ref{app:params}.} Then, we forecast the model~\eqref{eq:compact_etas}  to statistically stationary conditions using the $L$ sets of parameters, and we take from each time series a sample of length $t_\mathrm{tr}$, with this, we have the model estimates $\matr{Q}_\mathrm{u}\in\mathbb{R}^{L\times N_q\times N_\mathrm{tr}}$.  
Third, we correlate each $\matr{Q}_{\mathrm{u},l}$ with the data by selecting the time lag $\varkappa\geq0$ which minimizes their normalized root mean square error $(\mathrm{RMS})$ within the first 0.01~s (approximately 10 periods of the acoustic signal), i.e., for each $\matr{Q}_l$, the time lag is $\varkappa_l = \{\varkappa\;  s.t. \;\min{\mathrm{RMS}(\matr{Q}_{\mathrm{u},l}(t-\varkappa), \matr{D}(t))} \}$  with $t\in[0, 0.01]$~s. 
Once the $L$ model estimates are aligned to the observations such that the RMS is minimized in the initial 10 periods of oscillation, we obtain the training model bias and innovations as
\begin{equation}
    \matr{U}_l = \left[\matr{B}_{\mathrm{u},l};\matr{I}_{\mathrm{u},l}\right]= \left[\matr{D}^{\dagger}_\mathrm{u} - \matr{Q}_{\mathrm{u},l}; \matr{D}_\mathrm{u} - \matr{Q}_{\mathrm{u},l}\right], \quad \mathrm{for } \; l = 0, \dots, L-1  
\end{equation}
If our initial guess in the parameters is well defined, the training bias dataset $\matr{B}_{\mathrm{u}}$ have small norm. However, during assimilation the system can go through different states and parameter combinations, which may have a large norm model bias. Because the echo state network is only trained on the correlated signals, it may not be flexible to estimate the bias throughout the data assimilation process~\citet{liang_machine_2023}. 
Therefore, fourth, we apply data augmentation~\citep{goodfellow2016deep} to increase the robustness of the ESN and improve the network adaptive in a real-time assimilation framework. We increase the training set by adding the bias and innovations resulting from mid-correlated signals to the training set. We select the mid-correlation  time lag as the average between the best lag $\varkappa$ and the worst time lag. 
The total number of training time series in the proposed training method is $2L$. 
The training parameters used in this work are summarized in \ref{app:params}.
\begin{algorithm}
\caption{Training dataset generation}
\label{alg:train_dataset}
\begin{algorithmic}[1]
\State ${\matr{D}_\mathrm{u}}\gets$ \Call{Get observations}{observation time = $t_\mathrm{tr}$}
\State ${\matr{D}^\dagger_\mathrm{u}}\gets$ \Call{Estimate truth}{$\matr{D}_\mathrm{u}$}
\State ${\matr{U}}\gets${~[~]}
\For{$l=0 \;\mathbf{to}\; L-1$}
\State $\vect{\alpha}_l \gets \mathcal{U}(\vect{\alpha}^0(1- \sigma_L), \vect{\alpha}^0(1+ \sigma_L))$
\State $\matr{Q} \gets $  \Call{Forecast model}{$\vect{\alpha}_l, t_\mathrm{tr}$}
\State $\matr{Q}_{\mathrm{u}, l} \gets$ \Call{Correlate}{${\matr{D}_\mathrm{u}}, \matr{Q}$ } \Comment{Minimize RMS}
\State $\matr{U} \gets$ \Call{Append}{$[\matr{D}_\mathrm{u} -\matr{Q}_{\mathrm{u}, l};  \matr{D}_\mathrm{u}^\dagger -\matr{Q}_{\mathrm{u}, l}]$ }
\State $\matr{Q}^\mathrm{aug}_{\mathrm{u}, l} \gets$ \Call{Mid-correlate} {${\matr{D}_\mathrm{u}}, \matr{Q}$ }\Comment{Data augmentation}
\State $\matr{U} \gets$ \Call{Append}{$[\matr{D}_\mathrm{u} -\matr{Q}^\mathrm{aug}_{\mathrm{u}, l};  \matr{D}_\mathrm{u}^\dagger -\matr{Q}^\mathrm{aug}_{\mathrm{u}, l}]$ }
\EndFor
\end{algorithmic}
\end{algorithm}

\section{Simulations' parameters}\label{app:params}

This appendix summarizes the parameters used for the data assimilation and for training the echo state network. {The model parameters' ranges listed in Tab.~\ref{tab:simulaitons_params} are used to initialize the ensemble in the data assimilation algorithm, and to create the $L$-initial guesses for training the ESN (see \ref{app:ESN_train_data}).}
\begin{table}
\centering
\renewcommand{\arraystretch}{1.1} 
\caption{Parameters used in the simulations.}
\label{tab:simulaitons_params}
\newcommand{\one}[0]{$^\mathrm{\color{blue}a}$}
\newcommand{\three}[0]{$^\mathrm{\color{blue}b}$}
\newcommand{\mycline}[0]{\noalign{\vspace{1ex}}\cline{2-10}\noalign{\vspace{1ex}}}
\vspace{.5em}
\begin{tabular}{clr@{~}llr@{~}llr@{~}l}
\toprule
Assimilation  & $\Delta t_d$ & $35\Delta t$ & s & $N_q$ & 4 &  & $\epsilon_d$ & 0.1 &  \\
 & $\Delta t$ & ${51200}^{-1}$ & s  & $m$ & 20 & & Inflation  & 1.001 &   \\
\mycline
Model  & $\nu$\one & $[-10, 30]$ & Hz & $c_2\beta$\one & $[10, 50]$ & Hz & $\omega/(2\pi)$\one  & $[1090, 1095]$ & Hz \\
& $\Theta_\epsilon$\one & $[0.4, 0.6]$ & rad & $\kappa\E{4}$\one & $[1, 2]$ & Hz \\
& $\Theta_\beta$\one & $[0.5, 0.7]$ & rad & $\epsilon\E{3}$\one & $[5, 8]$&  \\
\mycline
ESN training& $\Delta t_\mathrm{ESN}$ & $5 \Delta t$ & s & $L$ & 50 &  & $\lambda$\three & $\{10^{-12}, 10^{-9}\}$ &  \\
 & $N_r$ & 50 &  & {Connectivity} & 3 &  & $\sigma_\mathrm{in}$\three & $[10^{-5},1]$ &  \\
 & $N_\mathrm{wash}$ & 10 &  & $t_\mathrm{tr}$ & 0.167 & s & $\rho$\three & $[0.5, 1.0]$&  \\
 & $N_\mathrm{folds}$ & 4 &  & $t_\mathrm{validate}$ & 0.020 & s &  &  & \\
\hline
\multicolumn{10}{l}{\footnotesize \one Indicates that the parameters are initialized within the given range.}  \\
\multicolumn{10}{l}{\footnotesize \three Indicates that the parameters are optimized in the given range. } 
\end{tabular}
\end{table}

\section{Inferred thermoacoustic parameters}\label{app:inferred_params}
Table~\ref{tab:all_params} details the parameters shown in Figures~\ref{fig:nu_beta_ERs}-\ref{fig:rest_of_params}. 
The reported parameters from the real-time digital twin (DT) are those with minimum RMS. 
\begin{table}
\centering
\caption{Thermoacoustic parameters inferred by the minimum-RMS solution (see Figure~\ref{fig:nu_beta_ERs}) of the digital twin (DT) and the bias-unregularized filter (EnKF) compared to the parameters identified  by \citet{indlekofer_noiray_spontaneous_2022} (ref.).
    }
    \label{tab:all_params}
    \begin{tabular}{cl@{~}crrrrrr}
    \toprule
         &  & & \multicolumn{4}{c}{Equivalence ratio} \\ 
         Param. & Units & Method & \multicolumn{4}{c}{\hrulefill}\\
         & & & \multicolumn{1}{c}{$\Phi=0.4875$} & \multicolumn{1}{c}{$\Phi=0.5125$}&\multicolumn{1}{c}{$\Phi=0.5375$}& \multicolumn{1}{c}{$\Phi=0.5625$} & \\ \midrule
         & &  ref. & \multicolumn{1}{c}{ $-22.43$}&  \multicolumn{1}{c}{ $-6.58$} &  \multicolumn{1}{c}{ $9.26$} & \multicolumn{1}{c}{ $25.11$} & \\
         $\nu$ & [Hz] & DT & $-18.48\pm3.76$ & $-4.84\pm 2.09$ & $12.03\pm3.21$ &  $25.13\pm 3.15$& \\
         & & EnKF & $-1.31 \pm 2.86$ & $-32.09 \pm 3.40$ & $4.58 \pm 2.57$ &  $24.42 \pm 2.02$ & \\ \midrule
         & &  ref. &  \multicolumn{1}{c}{ $17.65$} &  \multicolumn{1}{c}{ $24.11$} &  \multicolumn{1}{c}{ $30.57$} &  \multicolumn{1}{c}{ $37.02$} & \\
         $c_2\beta$ & [Hz] &DT & $30.64 \pm 8.51$ & $42.99 \pm 6.86$ & $46.84 \pm 7.18$ & $14.97\pm 3.97$ & \\
         & & EnKF & $27.77 \pm 5.14$ & $42.56 \pm 9.64$ & $49.55 \pm 5.38$ & $21.07 \pm 1.76$ & \\\midrule
         & &  ref. &  \multicolumn{1}{c}{ $1.20$}&  \multicolumn{1}{c}{ $1.20$} &  \multicolumn{1}{c}{ $1.20$} &  \multicolumn{1}{c}{ $1.20$} & \\
         $\kappa\E{4}$ & [Pa$^{-2}$s$^{-1}$] & DT& $1.55 \pm 0.20$ & $1.79\pm 0.19$ & $1.62\pm 0.16$ &  $1.10\pm0.12$& \\
         & & EnKF & $1.59 \pm 0.17$ & $1.30 \pm 0.21$ & $1.23 \pm 0.13$ & $1.19 \pm 0.09$ & \\\midrule
         & &  ref. & \multicolumn{1}{c}{ $1090.00$}& \multicolumn{1}{c}{ $1090.00$} & \multicolumn{1}{c}{ $1090.00$} & \multicolumn{1}{c}{ $1090.00$} & \\
         $\omega / (2\pi)$ & [rad~s$^{-1}$] & DT& $1092.36\pm 0.56$ & $1085.29\pm 0.20$ & $1089.05 \pm 0.15$ &  $1092.32 \pm 0.14$& \\
         & & EnKF & $1091.78 \pm 0.38$ & $1087.92 \pm 0.28$ & $1088.98 \pm 0.11$ &  $1092.19 \pm 0.04$ & \\\midrule
         & &  ref. & \multicolumn{1}{c}{ $2.30$} & \multicolumn{1}{c}{ $2.30$} & \multicolumn{1}{c}{ $2.30$} & \multicolumn{1}{c}{ $2.30$} & \\
         $\epsilon\E{3}$  & [~-~] &DT & $5.86\pm 0.57$ & $2.71 \pm 0.71$ & $5.48 \pm 0.55$ & $7.74\pm 0.52$ & \\
         & & EnKF & $6.34 \pm 0.38$ & $8.13 \pm 0.56$ & $5.66 \pm 0.41$ & $8.63 \pm 0.17$ & \\\midrule
         & &  ref. & \multicolumn{1}{c}{ $36.10$} & \multicolumn{1}{c}{ $36.10$} & \multicolumn{1}{c}{ $36.10$} & \multicolumn{1}{c}{ $36.10$} & \\
         $\Theta_\beta$ & [~$^\circ$~] & DT & $32.87\pm 1.83$ & $31.93\pm 1.84$ &  $29.99 \pm 0.65$& $35.50 \pm1.99$ & \\
         & & EnKF & $30.63 \pm 1.35$ & $34.07 \pm 1.60$ & $32.56 \pm 0.89$ & $31.64 \pm 1.04$ & \\\midrule
         & &  ref. & \multicolumn{1}{c}{ $37.82$} & \multicolumn{1}{c}{ $37.82$} & \multicolumn{1}{c}{ $37.82$} & \multicolumn{1}{c}{ $37.82$} & \\
         $\Theta_\epsilon$ & [~$^\circ$~]  & DT& $28.30 \pm 2.13$ & $21.72\pm 1.49$ & $30.05 \pm 0.62$ &   $28.87 \pm1.18$ &\\
         & & EnKF & $28.57 \pm 1.44$ & $26.03 \pm 1.87$ & $26.72 \pm 0.67$ &  $37.84 \pm 0.42$ & \\
    \end{tabular}
\end{table}


\end{document}